\documentclass[a4paper,11pt]{article}

\pdfoutput=1 

\usepackage{jheppub}  
\usepackage{simpler-wick}
\usepackage[T1]{fontenc} 
\usepackage{physics}
\usepackage{graphicx}

\usepackage{amsmath}
\usepackage{commath}
\usepackage{nccmath}
\usepackage{subcaption}
\usepackage{wrapfig,blindtext}
\usepackage{tcolorbox}
\usepackage{bm}
\usepackage{amsmath}
\usepackage{IEEEtrantools}
\usepackage{float}
\usepackage{hyperref}
\hypersetup{colorlinks,linkcolor={magenta},citecolor={magenta},urlcolor={magenta}}  
\usepackage{longtable}
\usepackage{pifont}
\usepackage{wasysym}
\usepackage{amssymb}
\usepackage{layout}

\usepackage[title]{appendix}

\def\beq{\begin{equation}}
\def\eeq{\end{equation}}
\def\bea{\begin{eqnarray}}
\def\eea{\end{eqnarray}}

\def\bit{\begin{itemize}}
\def\eit{\end{itemize}}

\def\baa{\begin{array}}
\def\eaa{\end{array}}

\def\simgt{\mathrel{\lower2.5pt\vbox{\lineskip=0pt\baselineskip=0pt
           \hbox{$>$}\hbox{$\sim$}}}}
\def\simlt{\mathrel{\lower2.5pt\vbox{\lineskip=0pt\baselineskip=0pt
           \hbox{$<$}\hbox{$\sim$}}}}

\def\bfc{\begin{figure}\begin{center}}
\def\efc{\end{center}\end{figure}}

\usepackage[top=1.8in,bottom=0.5in,left=1.9in,right=0.0in]{geometry}

\graphicspath{ {./image/} }

\title{\boldmath Mixed Graviton and Scalar Bispectra in the EFT of Inflation: Soft Limits and Boostless Bootstrap}

\author[a,1]{Diptimoy Ghosh}
\author[a,2]{Kushan Panchal}
\author[a,3]{Farman Ullah}

\affiliation[a]{Indian Institute of Science Education and Research Pune, India}

\emailAdd{diptimoy.ghosh@iiserpune.ac.in}
\emailAdd{kushan.panchal@students.iiserpune.ac.in}
\emailAdd{farman.ullah@students.iiserpune.ac.in}

\abstract{Boostless Bootstrap techniques  have been applied by many in the literature to compute pure scalar and graviton correlators. In this paper, we focus primarily on mixed graviton and scalar correlators. We start by developing an EFT of Inflation (EFToI) with some general assumptions, clarifying various subtleties related to power counting. 
We verify explicitly the soft limits for mixed correlators, showing how they are satisfied for higher derivative operators beyond the Maldacena action. We clarify some confusion in the literature related to the soft limits for operators that modify the power spectra of gravitons or scalars.  We then proceed to apply the boostless bootstrap rules to operators that do not modify the power spectra. Towards the end, we give a prescription that gives correlators for states that are Bogolyubov transforms of the Bunch-Davies vacuum, directly once we have the correlator for the Bunch-Davies vacuum. This enables us to bypass complicated in-in calculations for Bogolyubov states.
}

\begin{document}

\maketitle
\newpage

\section{Introduction}

It is believed that the structure in our universe was seeded by
quantum mechanical fluctuations generated during an epoch of
exponential expansion known as inflation \cite{STAROBINSKY198099,Guth:1980zm,LINDE1982389}.
During this period, these quantum fluctuations were stretched to
super-horizon distances with amplitudes freezing post horizon exit \cite{starobinsky:1979ty,Mukhanov:1981xt}.
Inflation
generates correlations between these fluctuations which seed the
late-time cosmological observables such as temperature correlations
on the CMB.  Although current observational reach is limited
to deducing the scalar power spectrum and its tilt \cite{Hazra_2014,Huang_2015,Planck:2018vyg}, one can expect
that in the future, higher point scalar correlators, as well as correlators
involving spinning fields such as the graviton, will also be measured. \\

There are a wide variety of models for Inflation \cite{Tong:2004dx,Arkani_Hamed_2004} and one can
compute observables for each of them. However, one can adopt a more
general (model-independent) approach by constructing an Effective
Field Theory of Inflation (EFToI) \cite{Cheung_2008,Weinberg_2008,Piazza_2013} consistent with symmetries.
This allows us to go beyond the minimally coupled canonical single
field slow-roll inflation \cite{Maldacena_2003}.
In this paper, we consider higher derivative operators which contribute to the
mixed three-point correlation functions (i.e. $\langle \zeta\zeta\gamma \rangle$,
$\langle \gamma\gamma\zeta \rangle$, which we compute using the in-in formalism).
It is important that the EFT respects  soft
limits \cite{Maldacena_2003,Creminelli_2012,Bordin_2020,Cabass_2022} provided that we do not violate any of the assumptions implicit in their derivations. We clearly state under what assumptions these theorems are expected to hold and explicitly check (for mixed correlators) that they are satisfied for operators that modify both scalar as well as tensor power spectra. This  provides important consistency checks for our calculations.  \\

It is also interesting to bootstrap these correlators from a pure boundary
perspective without referring to the bulk evolution.
This approach has a lot of advantages, for instance,
it was shown in \cite{Creminelli_2014,Bordin_2017} that using certain field redefinitions
(that vanish at the boundary)  some operators can be removed from the EFT without
affecting the late-time correlators. This redundancy, by construction, is not
present once we have a purely boundary perspective and
therefore can potentially lead to a lot of simplifications.
 There is a large amount of literature  on both the conformal/pure
 de-sitter bootstrap \cite{Mata_2013,Kundu_2016,https://doi.org/10.48550/arxiv.1811.00024,https://doi.org/10.48550/arxiv.2203.08121} as
 well as the ``boostless bootstrap'' approach, the latter being more recent
 \cite{Pajer_2021,Cabass:2021fnw,Goodhew_2021,Jazayeri_2021}. The boostless bootstrap program focuses
 on properties such as the analytical structure of the correlators, soft limits etc.
 The analytical structure of correlators on its own gives a lot of information such
 as the initial state \cite{Green_2020,Green_2022,https://doi.org/10.48550/arxiv.2207.06430} and the
 flat space amplitude \cite{Bordin_2020,https://doi.org/10.48550/arxiv.1811.00024}
 corresponding to the interaction. This technique has had considerable success for
 pure graviton correlators  (sometimes in conjunction with tools like spinor helicity
 formalism and results related to parity and the cosmological
 optical theorem \cite{Maldacena_2011,https://doi.org/10.48550/arxiv.2210.02907,Cabass:2021fnw}).
 In this paper, we follow the rules entailed in \cite{Pajer_2021} to bootstrap
 three-point mixed correlators arising from operators in our EFToI that do not
 modify power spectra, and check to what extent the Boostless Bootstrap works. In doing so, we separately consider local and non-local interactions.\\

 Finally, we give a prescription for obtaining the correlators for states that are Bogolyubov tranformations (BT) of Bunch-Davies (BD) vacuum. As an example, we calculate the correlators for $\alpha$- vacua (which
 is the most general family of vacuum states consistent with de Sitter isometries \cite{Allen:1985ux})
 from the answers for BD vacuum. This offers significant
 computational benefit since one does not have to repeat the cumbersome in-in calculation for BT states.
 

\section{Notations and Conventions}
For all the calculations that follow, the form of the metric (in unitary gauge) is given by the standard ADM formulation \cite{Arnowitt:1962hi}:

\begin{flalign}
ds^2&=-N^2dt^2+ g_{ij}(dx^i + \widetilde{N^i}dt)(dx^j + \widetilde{N^j}dt)
\end{flalign}
where we see that:
\begin{flalign}
g_{00}=-N^2+ &g_{ij} \widetilde{N^{i}}\widetilde{N^{j}} 
\end{flalign}
and we define:
\begin{flalign}
&g^{0i}=N^i \, , ~~~~g_{ij}=a^2e^{2\zeta}e^{\gamma_{ij}} \, ,~~~~\partial_{i}\gamma_{ij}=0 \, , ~~~~~\gamma_{ii}=0    
\end{flalign}

where $a$ is the scale factor and $\frac{\dot{a}}{a}=H$ is the Hubble constant. $\gamma_{ij}$ is the  gauge-invariant tensor perturbation and $\zeta=\psi+H\frac{\delta\phi}{\bar{\dot{\phi}}}$ is the gauge-invariant scalar perturbation. It is important to note that the quantity that actually appears in $g_{ij}$ is $\psi$ but since in unitary gauge $\delta\phi=0$, we have $\psi|_{\delta\phi=0}=\zeta$. For $\gamma_{ij}$, we denote the two helicities by $h$ and the corresponding polarization tensors by $e^{h}_{ij}$. We follow the normalization convention: $e^{h_1}_{ij}(\bm{k})e^{h_2}_{ij}(\bm{-k})=2\delta _{h_1h_2}$ . In terms of $H$ and the background inflaton field $\phi(t)$,the slow roll parameters are defined as:
\begin{align}
    \epsilon= -\frac{\dot{H}}{H^2} =\frac{\dot{\phi}^2}{2 M_{pl}^2 H^2}~~~~~~~~~~\eta= -\frac{\ddot{\phi}}{\dot{\phi}H}+\frac{\dot{\phi}^2}{2 M_{pl}^2 H^2}     
\end{align}

\section{EFToI: Explicit check of soft theorems}\label{sec-3}
In this section, we will explicitly check the validity of soft theorems for some of our EFT operators but first, we  quickly review the key assumptions implicit in their derivations. It was shown in \cite{Maldacena_2003,Creminelli:2004yq,Cheung:2007sv,Creminelli_2012} that every single field inflation model must satisfy consistency conditions relating three-point functions (in the squeezed limit) to two-point functions. 
\begin{align}
    \lim_{\bm{k_1}\rightarrow 0}\langle\zeta_{\bm{k_1}}\zeta_{\bm{k_2}}\zeta_{\bm{
    k_3}}\rangle=-(n_{s}-1)\langle\zeta_{\bm{k_1}}\zeta_{-\bm{k_1}}\rangle\langle\zeta_{\bm{k_2}}\zeta_{-\bm{k_2}}\rangle\left(1  {+ \mathcal{O}(k^{2}_1/k^{2}_2)}\right)\label{sl-2}\\
     \lim_{\bm{k_1}\rightarrow 0}\langle\zeta_{\bm{k_1}}\gamma^{h}_{\bm{k_2}}\gamma^{h'}_{\bm{k_3}}\rangle=-n_{t}\langle\zeta_{\bm{k_1}}\zeta_{-\bm{k_1}}\rangle\langle\gamma^{h}_{\bm{k_2}}\gamma^{h'}_{-\bm{k_2}}\rangle\left(1 {+ \mathcal{O}(k^{2}_1/k^{2}_2)}\right)\label{sl-1}\\
     \lim_{\bm{k_1}\rightarrow 0}\langle\gamma^{h}_{\bm{k_1}}\zeta_{\bm{k_2}}\zeta_{\bm{k_3}}\rangle=-\langle\gamma^{h}_{\bm{k_1}}\gamma^{h}_{-\bm{k_1}}\rangle\epsilon^{h}_{ij}k_2^{i}k_2^{j}\frac{\partial}{\partial k_2^{2}}\langle\zeta_{\bm{k_2}}\zeta_{-\bm{k_2}}\rangle \left(1 {+ \mathcal{O}(k^{2}_1/k^{2}_2)} \right)\label{sl-0}
\end{align}
where, $\zeta$ and $\gamma$ are defined in the Appendix \ref{apA}, and $n_{s}$, $n_t$ are the scalar and tensor spectral tilts defined as 
\begin{align}
    n_s-1=\frac{d\ln \left(k^{3}\langle\zeta_{-\vec{k}}\zeta_{\vec{k}}\rangle\right)}{d\ln k}\\
        n_t=\frac{d\ln \left(k^{3}\langle\gamma_{-\vec{k}}\gamma_{\vec{k}}\rangle\right)}{d\ln k}
\end{align}
For the leading order (two derivatives) minimally coupled canonical action, we have $n_s-1=2(\eta-3\epsilon)$ and $n_t=-2\epsilon$ \cite{Maldacena_2003}. Besides requiring only a single degree of freedom during inflation and the condition that the curvature perturbation does not grow outside the horizon (not true for non-attractor models \cite{Kinney:2005vj,Huang:2013oya,Chen:2013eea,Martin:2012pe,Namjoo:2012aa}, see \cite{Finelli:2017fml,Finelli:2018upr} for modified soft theorems in shift symmetric non-attractor models), these relations also rely on the fact that there is a negligible contribution to the three-point function when all the modes are within the horizon. In fact, these theorems assume that the contribution to the correlators only starts becoming sizable when the long mode has left the horizon and has already frozen  acting as a classical background. This is always the case when the initial state is the BD vacuum. The BD three-point correlators only have a total energy pole which comes from integrals of the form
\begin{equation}
    \langle\zeta_{\vec{k_1}}\zeta_{\vec{k_2}}\zeta_{\vec{k_3}}\rangle\sim\int_{-\infty}^{0}\tau^{m}e^{iK\tau}d\tau
\end{equation}
where, $K=k_1+k_2+k_3$. To compute such an integral, a regularisation scheme is required. This integral can be regularized by taking $\tau\rightarrow\tau(1-i\epsilon)$, and therefore, damping the contribution in the far past. It only starts giving appreciable contribution once $K\tau\sim-1$ which in the equilateral case ($k_1=k_2=k_3$) corresponds to the epoch of horizon crossing for all the modes and in the squeezed case ($k_1\ll k_2\approx k_3$, relevant for soft theorems) corresponds to epoch of horizon crossing for the short modes since $K\sim k_s$. Therefore, in the squeezed limit the in-in contribution remains negligible till the epoch of the horizon crossing for the short modes. By this time the long mode has already frozen and therefore, the soft theorems must hold. There exist cases where the soft theorems are violated even in single field models \cite{Namjoo:2012aa,Chen:2013aj,Cai:2018dkf,https://doi.org/10.48550/arxiv.2207.06430}, for instance, if one starts in an excited initial state, for e.g., $\alpha $-vacuum \cite{Allen:1985ux} then apart from total energy pole, one also has a $\left(k_1+k_2-k_3\right)$ kind of pole structure arising from mode-mixing \cite{Holman_2008}. Therefore the integral contains terms,
\begin{equation}
      \langle\zeta_{\vec{k_1}}\zeta_{\vec{k_2}}\zeta_{\vec{k_3}}\rangle\sim\int_{-\infty}^{0}\tau^{m}e^{i\left(k_1+k_2-k_3\right)\tau}d\tau
\end{equation}
which, in the squeezed limit ($k_1\ll k_2\approx k_3$) starts giving appreciable contribution when $k_1\tau\sim-1$ i.e when the long mode is near horizon exit. Therefore, the soft theorem proof does not hold in this case.

\subsection{EFToI}
The Effective field theory of Inflation is an attempt to unify all models of inflation by constructing the most general low energy action consistent with symmetries. The action is written in terms of fluctuations themselves. In cosmological perturbation theory (See \cite{Mukhanov:1990me} for a nice review) one always decomposes any relevant quantity (e.g., $\phi(x)$) in the following manner
\begin{equation}
    \phi(\vec{x},t)=\bar{\phi}(t)+\delta\phi(\vec{x},t)
\end{equation}  
where $\bar{\phi}(t)$ is the solution of the following background EOM for the scalar field (FRW coordinates)
\begin{equation}
    \ddot{\phi}(t)+3H\dot{\phi}(t)+V'=0
\end{equation}
No matter what coordinates one works with, the split between background and perturbation will always have the same time-dependent (no spatial dependence) background function. The reason for such a split is the following: our universe on very large scales looks very homogeneous (no spatial dependence) and isotropic to co-moving observers. This has been verified by the Large scale structure data and the CMB temperature profile \cite{Planck:2018vyg}. Of course, this is only an approximate statement and there are small fluctuations present on top of this background. Since we are interested in studying these fluctuations we always split any relevant quantity in any coordinates in terms of an FRW background solution (homogeneous and isotropic) and perturbations about it. If one thinks in terms of a background spacetime then this means that there is a preferred choice of the time coordinate (slicing) in which the background metric has no spatial dependence. For the inflationary background, this slicing corresponds to slices of constant inflaton which act as a  physical clock (with values of this field keeping time). This is the reason why sometimes single-field inflation is also referred to as single-clock inflation.\\
\noindent As mentioned above, we are finally interested in a theory of fluctuations about this time-dependent background and therefore time diffeomorphism while still being a symmetry is now non-linearly realised on the fluctuations i.e they are spontaneously broken. One can then choose a gauge to write the EFT in the so-called unitary gauge which has no inflaton fluctuations $\delta \phi(\Vec{x},t)=0$. In this gauge, all fluctuations go into the metric. The transformation rule for $\delta\phi(\vec{x},t)$ under $x^{\mu}\rightarrow x^{\mu}+\epsilon^{\mu}$,
\begin{equation}
    \delta\tilde{\phi}(\tilde{x})=\delta\phi(x)-\epsilon^{0}\dot{\bar{\phi}}(t)
\end{equation}
indicates that unitary gauge just fixes the time diffs and therefore, the EFT will only involve terms which are invariant under spatial diffeomorphisms. We are allowed to write full diff invariant terms, e.g.,$^{(4)}\!R_{\mu\nu}^{(4)}\!R^{\mu\nu}$, terms with free upper time indices, e.g., $\delta g^{00}$ (terms with lower free time indices are not invariant under spatial diffs) and terms describing the slicing. Hence we have:
\begin{equation}
    S_{EFT}=\int_{}{} ~d^4 x \sqrt{-g} \mathcal{L}(^{(4)}\!R_{\mu \nu \sigma \rho},^{(3)}\!R_{ijkl}, \nabla_{\mu}, \delta K_{ij},g^{00},\partial_t,...)
\end{equation}
So far we have identified the correct degrees of freedom for the action. Now, like for any sensible EFT, one must identify the correct expansion parameter(s) to organise the terms. The EFToI is an expansion in the number of metric perturbations and the number of derivatives on the metric perturbations which (prior to any canonical normalization) have dimension 0. We start by splitting the action into three parts,
\begin{equation}
    S_{EFT}=S_{0}+S_{2}+S_{\geq3}
\end{equation}
where,
\begin{align}
\begin{split}
    ~~~~~~~~~~~~~~~~~~~~~~~~\\
    S_{2}=\int_{}{}~d^4x ~\sqrt{-g }M^2_{pl}\left ( m_1\left(\delta K^i_j \delta K^j_i\right) + m_2(\delta K)^2+m_3\left(^{(3)}\!R\delta g^{00}\right)+M_2^{2} (\delta g^{00})^2+M_3\left(\delta g^{00}\right)\delta K\right.\\ \left.+\frac{1}{M_4}{^{(3)}\!R_{ij}}\delta K^{ij}+\frac{1}{M_5}{ ^{(3)}\!R }\delta K+
   \frac{1}{M^{2}_6}{^{(3)}\!R}^{2}+\frac{1}{M^{2}_7}{^{(3)}\!R}^{2}_{ij}+\frac{1}{M^{2}_8}{^{(3)}\!R}^{2}_{ijkl}\right)+....\\
   \end{split}\label{eft}
   \end{align}
where, $S_{n}$ has  terms which start from $n^{th}$ order in perturbations. Here, $S_{0}$ is Einstein-Hilbert action coupled with a scalar field with $\lambda(t)=-3H^2 + \dot{H}$ and $c(t)=-\dot{H}$ so that the background/zeroth-order equations of motion are satisfied. The dots in the above equations indicate the same terms but with more derivatives. $\{m_{i}\}$ have mass dimension zero whereas $\{M_{i}\}$ have mass dimension one. We have not made any assumption about how the EFT operators are generated from the UV, therefore every operator comes with a different Mass scale, $M_{i}$. It is important to keep in mind that finally, we want a theory in terms of the scalar fluctuations $\zeta$ and the tensor fluctuation $\gamma_{ij}$. The EFT derivative power counting is clear only in terms of these variables and to illustrate this consider the operator $\partial_{i}\delta g^{00}\partial^{i}\delta g^{00}$. This naively looks like a leading two-derivative term and therefore, is not suppressed or enhanced by any mass scale. However, in terms of $\zeta$, depending on the constraint solution, it can actually generate higher derivative terms. Therefore, some terms in the relevant part of Lagrangian might actually turn out to be irrelevant. Since we are ignorant about the short distance physics i.e. $k \gg H$ and are interested in the modes which have exited the horizon and have left an imprint on the CMB, the typical length/energy scale is $H$. Hence an expansion in $H/M$ is obtained for our EFToI. Assuming for concreteness that all dimensional coefficients are order unity and $H\ll M_{4,5,6,7}$, one can easily arrange the action as a perturbative series. For instance, the quadratic action, $S_2$, is
\begin{align}
    S_{2}\sim\left( m_1+ m_2+m_3+\frac{M_2^{2}}{H^{2}}+\frac{M_3}{H}+\frac{H}{M_4}+\frac{H^{2}}{M^{2}_5}+\frac{H^{2}}{M^{2}_6}+\frac{H^{2}}{M^{2}_7}\right)+....
   \end{align}
  For some explicit calculations, we take the action $S_0 + S_2$ and write only the operators which have at most 2 derivatives on the metric :
\begin{flalign}\label{2.4}
    S_0+ &S_2=\int_{}{}~d^4x ~\sqrt{-g }M^2_{pl}\left ( \frac{ ^{(4)}\!R}{2}+ m_1\delta K^i_j \delta K^j_i + m_2(\delta K)^2 + m_3 ^{(3)}\!R \delta g^{00}+ D\delta K -M^2_{pl}\lambda(t) -c(t)g^{00} \right.  &&\nonumber\\
    &\left. +M_1 g^{ij} \partial_i \delta g^{00}\partial_j \delta g^{00} +
     M^2_2 (\delta g^{00})^2 + M_3 \delta g^{00}\delta K\right)
\end{flalign}
 The leading (two derivatives) quadratic (scalar and tensor) actions  are given by \cite{Maldacena_2003} :
\begin{align}
S_{\zeta\zeta}=& \frac{1}{2}\int_{}{}d^4x ~\epsilon \left (a^{3} \dot{\zeta}^2 - a(\partial\zeta)^2 \right) \label{q-1}\\
S_{\gamma\gamma}=&\frac{1}{8}\int_{}{}d^4x~\left (a^{3} \dot{\gamma_{ij}}^2 - a(\partial\gamma_{ij})^2 \right)\label{q-2}
\end{align}
i.e the action derived for a canonical scalar field  minimally coupled to gravity. This is an element of the class of actions defined by equation (\ref{2.4}) where the action only contains the first three terms of $S_{0}$. In passing we point out that from (\ref{q-1}) and ($\ref{q-2}$) one can easily see that the existence of $\zeta$ is tied to the fact that inflation is quasi de Sitter ($\epsilon\neq0$), and such a variable does not exist in pure de Sitter ($\epsilon=0$). On the other hand, the tensor perturbation $\gamma_{ij}$ is also well defined in de Sitter. Coming back to (\ref{2.4}), there is a large number of unknown parameters but as usual, the EFT parameters have to be determined/constrained experimentally. Unfortunately in cosmology, the number of observables is very small. This is due to the fact that unlike in flat space EFTs, we do not have experimental control. Therefore, as we will show below, one can constrain or fix only a handful of parameters in the EFToI:
\begin{itemize}
    \item  If a mass term is generated for the scalar perturbations in the EFT ($c_1 \dot{\zeta}^2 - c_2 (\partial\zeta)^2-c_3\zeta^2$, where $m^{2}=c_3/c_1$ ) then from the observed tilt of the scalar spectrum one can show that $\frac{m}{H}\ll1$, where $m$ is the mass of the fluctuations. This can be easily inferred from the expression for power spectrum for a massive scalar in de-Sitter. 
    \begin{equation}
        P_{\zeta}(k)=\frac{H^{2}}{2k^{3}}\left(\frac{k}{aH}\right)^{3-2\nu}
    \end{equation}
    where, $\nu = \sqrt{\frac{9}{4}-\frac{m^{2}}{H^{2}}}$. Now, since we know (through observations) that the spectral tilt is very close to unity, therefore $\nu\sim\frac{3}{2}\implies\frac{m^{2}}{H^{2}}\ll1$. This allows us to strongly constrain the coefficients of operators which generate a mass term for $\zeta$. 
\item For operators which contribute both to the sound speed and non-gaussianities, we can use experimental bounds to deduce the bounds on $c_s$ or the operator coefficients. Further constraints on $c_s$ or the coefficients can be found by applying  the partial wave unitarity bound by  going to the flat gauge \cite{Cheung_2008,Baumann_2016}.
\end{itemize}
 To illustrate the points mentioned above, we take two examples of quadratic actions of the type mentioned in Equation (\ref{2.4}). We first solve for the ADM constraint variables  up to 1st order in $\zeta$ (See Appendix \ref{apA}). To simplify calculations, we take $m_2=-m_1$ \footnote{\textcolor{black}{This can be guaranteed by field redefinitions of $g_{ij}$ as shown in \cite{Bordin_2017}. However, the coefficients of other operators might change as well so $m_1, m_2$ may still enter the expressions for the constraint equations.}}. We then consider the cases:

\begin{itemize}
    \item \bm{$M_3=0, m_1,m_2 \neq 0$}. The canonical 2 derivative terms in the action are given by:
    \begin{equation}
        S_0+S_2= \int_{}{} M^2_{pl}\left( a^3 \left(\epsilon + 4\left(\frac{M_2}{H}\right)^2 \right)\dot{\zeta}^2- a \epsilon (\partial \zeta)^2 \right) 
    \end{equation}
i.e. the usual canonical action with a different sound speed. The partial wave unitarity bound in the flat space limit is satisfied in the limit $c_s \to 1$ \cite{Cheung_2008}
    , which gives $M_2 \ll \sqrt{\epsilon} H$. For this theory, which has no mass term, we have the spectral tilt:

    \begin{equation}
n_s-1= \frac{1}{H}\frac{d}{dt_*} \langle \zeta(\bm{k}) \zeta(\bm{-k})\rangle=  \frac{1}{H}\frac{d}{dt_*} \frac{H^2(t_*)}{4\epsilon(t_*) M_{pl}^2 c_s(t_*) k^3}  =2\eta -6\epsilon + 4 c_s ^2  \left(\frac{M_2}{H}\right)^2 (\frac{\eta}{\epsilon}-1)\label{ns-1}
\end{equation}
where the asterisk denotes the time of horizon crossing. Because of the $c_s$ constraint, the spectral tilt is already small. As was shown in \cite{Cheung_2008}, a small speed of sound ($c_{s}\ll1)$ also implies large interactions (non-gaussianities). These are derivative interactions (e.g., $\dot{\pi}^{3}$) and they naturally produce equilateral non-gaussianity since, due to derivatives their contribution in the squeezed limit is negligible.
There are experimental constraints on $c_s$ from bounds on equilateral non-gaussianity, $f^{equil}_{NL}$,
\begin{equation}
    c_s\gg0.028
\end{equation}

    \item $\bm{M_3 \neq 0, m_1=m_2=0, M_2=0}$. Since calculating the action for arbitrarily large values of $M_3$ is difficult, we take the case where $g=M_3/ H\ll1$. For this we get upto 1st order in $\epsilon$ and $g$,
    
\textcolor{black}{\begin{equation}
S_0+S_2=\int_{}{} M^2_{pl}\left( a^3\epsilon\dot{\zeta}^2- a\left(\epsilon-2g \right) (\partial \zeta)^2 \right) 
\end{equation}}
 Again, taking into account that mass terms only start appearing at $\mathcal{O}(g^2)$, the spectral tilt is small and we won't explicitly calculate it here.
 Going to the Stueckelberg gauge \cite{Cheung_2008}, we can write the lagrangian in the flat space limit, in terms of the Stueckelberg boson $\pi$ as :

 \begin{equation}
     \mathcal{L}_{\pi}= M^2_{pl} \epsilon H^2 \left(\dot{\pi}^2-(\partial \pi)^2 \left(1-\frac{M_3}{H\epsilon} \right) \right)-2\frac{M_3 M_{pl}^2c_s^3}{\left(\sqrt{M^2_{pl}\dot{H}}\right)^3} \partial^2\pi (\dot{\pi}^2+ (\partial \pi)^2) + ~~ (\text{$\epsilon$ -suppressed terms})
 \end{equation}
 This lagrangian tells us that we must have $M_3>0$ for $c_s<1$ and the partial wave unitarity bound for the tree level $\pi\pi$ amplitude gives: 
\begin{equation}
    \frac{M^2_3}{\epsilon^3 M^2_{pl}c_s^2} \left(c_s^4+4(1+c_s^2)^2\right)<\pi
\end{equation}
 where, in the line above, the $\pi$ is the numerical constant.
\end{itemize}
While all this is for the scalar action, the graviton case is simpler to deal with since the only two operators contributing to the two derivative quadratic EFT are $^{(3)}\!R$ and $\delta K_{ij} \delta K^{ij}$ which can be removed by suitable field re-definitions of $g_{\mu \nu}$ \cite{Creminelli_2014,Creminelli_2012}. We also note that one can write the EFT in some other gauge, for instance, the flat gauge where $\delta \phi \neq 0$ \cite{Maldacena_2003}. Doing calculations in flat gauge can lead to simplifications as it can sometimes directly give us an EFT ordered in $\epsilon $ \cite{Maldacena_2003}. One has to be careful while doing the calculations in this gauge and converting them to the unitary gauge answers. This is pointed out using an example in Appendix \ref{apE}.


\subsection{Explicit Checks}
In this section, we explicitly calculate and verify the soft limits mentioned above. We start with the scalar three-point function calculated for the action in \ref{2.4} with only $M_{2}\neq 0$ as a check for calculations with $c_s \neq 1$ while for the mixed correlators,  we take the canonical minimally coupled quadratic  action for simplicity of calculation. Although the scalar soft theorems have been explicitly checked \cite{Cheung:2007sv} the mixed correlator ones have not been explicitly checked for higher derivative EFToI operators and therefore, according to our knowledge, this is the first such an attempt.

\subsubsection{Soft limit for $\langle \zeta \zeta \zeta \rangle$}
The terms which contribute to the soft 3-point correlator $\langle \zeta(k_1) \zeta (k_2) \zeta (k_3) \rangle |{k_1 \to 0}$ are given by:

\begin{equation}
   S_0+S_2= \int_{}{} M_{pl}^2 \left[\left(\epsilon ^2 \zeta_c\dot{\zeta_c}^2 a^{3}  + \epsilon^2 \zeta_c (\partial \zeta_c)^2 a\right)  + 4\left(3\epsilon-2\eta\right)\left(\frac{M_2}{H}\right)^2\zeta_c \dot{\zeta_c}^2 a^{3} \right]\label{2.28}
\end{equation}
where the first two terms are from the Maldacena cubic action. It is important to note that the final Maldacena cubic action is derived after performing a field redefinition \cite{Maldacena_2003} $$\zeta=\zeta_c + \frac{1}{2}(2\epsilon-\eta) \zeta^2_c+\frac{\dot{\zeta_c}\zeta_c}{H}+....$$ 
which removes terms proportional to the equation of motion. Since $M_2$ also gives quadratic corrections to $\zeta$ action, the above field redefinition generates additional cubic terms. Therefore, the last term in \ref{2.28} is a combination of the cubic part generated by $M_2$ and the terms generated by the redefinition. The soft correlator is given by (where $n_s$ is given by \ref{ns-1}),

\begin{equation}
\begin{split}
    \langle \zeta(\bm{k_1})\zeta(\bm{k_2})\zeta(\bm{k_3})\rangle_{\bm{k_1}\to 0}= &\left(\frac{H^2}{4\epsilon c_s M_{pl}^2}\right)^2 \left( \frac{11}{2}\epsilon - 2\eta +  \frac{\epsilon}{2}c_s^2 +4 \frac{M^2_2}{H^2} c_s^2\left(\frac{3}{2}-\frac{\eta}{\epsilon} \right)  \right) \frac{1}{k_1^3}\frac{1}{k_2^3}\\
=&-(n_s-1) \langle \zeta(\bm{k_1}) \zeta(\bm{-k_1})\rangle \langle \zeta(\bm{k_2}) \zeta(\bm{-k_2}) \rangle
\end{split}
\end{equation}

\subsubsection{Soft limits for mixed correlators}
We classify the operators in the EFT into two classes, Cubic and Purely Cubic operators. Cubic operators are those which contribute to $\langle \zeta \zeta \rangle$ or $\langle \gamma \gamma\rangle$, as well as the mixed bispectra, while purely cubic operators contribute only to the latter. We take two cubic operators from \ref{eft} as examples: \\[0.2cm]

\bm{$(1)$}$\bm{\int  d^4{x}\sqrt{-g}\frac{M^2_{pl}}{M_5} {^{(3)}\!R}\delta K:}$\\ Since $\delta K$ contains both $\delta N$ and $N^{i}$, i.e. the ADM constraint variables, they get modified and are now given by:
\begin{flalign}
   \delta N=&\frac{\dot{\zeta}}{H}+\frac{2\partial^2 \zeta a^{-2}}{M_5 H} \\
    N^i =&\partial_i \left( -\frac{\zeta}{H}a^{-2}+ \epsilon \partial^{-2}\dot{\zeta}+ \frac{2\epsilon \zeta a^{-2}}{M_{5}}\right)
\end{flalign}

Note that the two equations above are valid only when $\frac{H}{M_{5}}\ll1$ i.e. they're 1st order expressions in $1/M_5$. Using this, the corrections to the quadratic and cubic actions for $\gamma \gamma \zeta$ are given up to first order in $1/M_5$ by:

\begin{flalign}
\mathcal{O}_{\zeta\zeta}=&\int_{}{} 4 \frac{M^2_{pl}}{M_5} a \partial^2 \zeta \left(\epsilon \dot{\zeta}- \frac{\partial^2 \zeta}{H}a^{-2} \right) &&\\ 
\mathcal{O_{\gamma\gamma\zeta}}=&\int_{}{} \frac{M^2_{pl}}{4M_5}\left( a \frac{\dot{\gamma_{ij}}\dot{\gamma_{ij}}} {H} \partial^2 \zeta- 2a^{-1} \frac{\partial_l \gamma_{ij}\partial_l\gamma_{ij} \partial^2 \zeta}{H} + \epsilon a\partial_l\gamma_{ij}\partial_l\gamma_{ij} \dot{\zeta}-2\epsilon\dot{\gamma_{ij}\partial_l\gamma_{ij}\partial_l \zeta}\right)&&
   \end{flalign}
   
The correction to the scalar power spectrum is given by,
    \begin{equation}
        \delta P_{\zeta}= -\frac{M^2_{pl}}{M_5}\frac{H^3}{4 \epsilon^2 M^4_{pl} k^3} (5-3\epsilon) 
    \end{equation}
For perturbation theory to work here, the quadratic correction $\delta P_{\zeta}$ must be small compared to $P_{\zeta}(k)=\frac{H^2}{4\epsilon M_{pl}^2 k^3}$. We henceforth assume that $\frac{ H}{M_5\epsilon} \ll1$ so that the enhancement to the power spectrum remains small. This also ensures that we can expand the action (as well as the ADM constraints) in powers of $M$ and the analysis above holds. In the soft limit, $\mathcal{O}_{\gamma \gamma \zeta_{k \to 0}}=0$ i.e its contribution to the mixed correlator in the soft limit vanishes at leading order. However, the following ``exchange diagram'' diagram gives a non-zero contribution:

\begin{figure}[H]
\includegraphics[width=10.5cm, height=5.3cm]{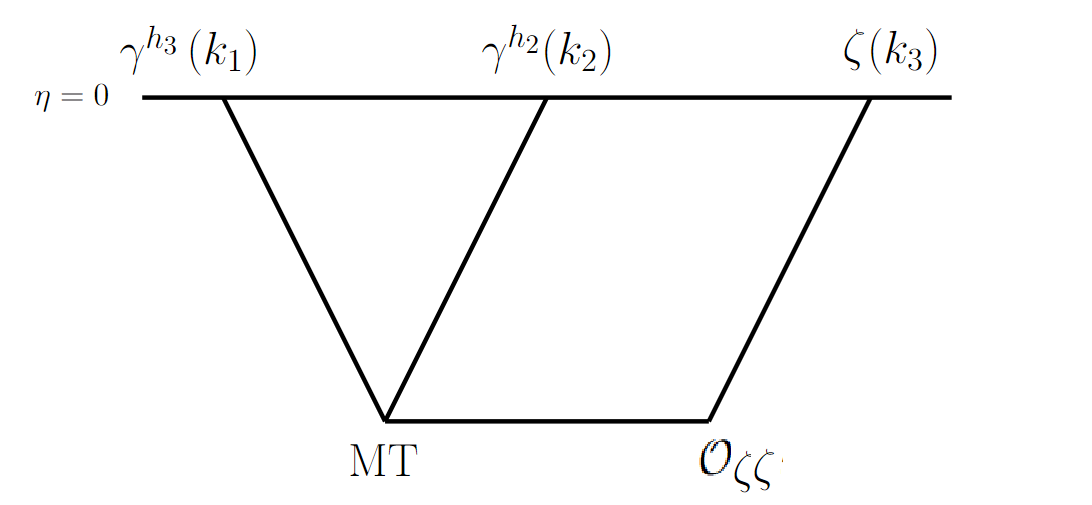}
\centering
\caption{The MT (Maldacena term) vertex is the usual cubic $\gamma\gamma \zeta$ vertex of the Maldacena action}
\end{figure}

where the Maldacena term refers to the mixed cubic vertex ($\zeta\gamma\gamma$) computed in \cite{Maldacena_2003}. This 3-point correlator is a sum of two terms,
\begin{equation}
    \langle \gamma(\bm{k_1})\gamma(\bm{k_2})\zeta(\bm{k_3}) \rangle_{\bm{k_3}\to 0}= 2\left(\langle \gamma(\bm{k_1})\gamma(\bm{k_2})\zeta(\bm{k_3}) \rangle_{LR, {\bm{k_3}\to 0}} -\langle \gamma(\bm{k_1})\gamma(\bm{k_2})\zeta(\bm{k_3}) \rangle_{RR, {\bm{k_3}\to 0}}\right)
\end{equation}
where $R$ and $L$ indicate the time and anti-time orderings of the interaction Hamiltonian respectively. For instance, RR means both vertices are time ordered (See \cite{Weinberg_2005} for more details on these notations and conventions). The $RR$ contribution is given by \footnote{The results given from here on do not include the divergent terms of the type: $\lim_{\eta \to 0} \frac{cos(k \eta)}{\eta}$. These are ignored while calculating contact diagrams since they contribute only to the imaginary part of the $L$ and $R$ vertices and so, they get cancelled. Here, however, these terms come from both the vertices of the ``exchange'' diagram and get multiplied, because of which they contribute to the real part. However, one can easily check that the contributions from $RR, RL, LR$ and $LL$ add up to $0$. }

\begin{flalign}
    &\langle \gamma(\bm{k_1})\gamma(\bm{k_2})\zeta(\bm{k_3}) \rangle_{RR},_{\bm{k_3}\to 0 }
    =\left(\frac{H^6}{M^6_{pl} k_1^3k_2^3 k_3^3 } \right) \left(\frac{M^2_{pl} }{8\epsilon H M_5}\right)e^{h_1}_{ij}(k_1)e^{h_2}_{ij}(k_2)\left[ k_1^2 k_2^2 \left[ \frac{5-3\epsilon}{4k_T}+\frac{(3-\epsilon)k_3}{k_T^2}-\right.\right.
    &&\nonumber\\ 
    &\left. \left. \frac{\epsilon k_3^2}{k_T^3} -\frac{5k_3^3}{k_T^4}-\frac{8k_3^4}{k_T^5}\right]
  -(\bm{k_1}\cdot \bm{k_2}) \left[\frac{5-3\epsilon}{4} k_T + \epsilon k_3 I -\frac{5-3\epsilon}{2} k_3 +(3-\epsilon)\frac{k_3^2}{k_T}-k_1k_2\left(\frac{5-3\epsilon}{4k_T} +\frac{5-3\epsilon}{2k_T^2}+ \right.\right. \right.&&\nonumber\\
    &\left.\left. \left. \frac{k_3^2}{k_T^3}-2\frac{k_3^3}{k_T^4}  \right)+(k_2k_3+k_1k_3)\left(\frac{5-3\epsilon}{4k_T}+ \frac{(1+\epsilon)k_3}{2k_T^2}+\frac{k_3^2}{k_T^3}+\frac{2k_3^3}{k_T^4}\right)+k_1k_2k_3 \left( \frac{5-3\epsilon}{4k_T^2}+  \frac{(1+\epsilon)k_3}{k_T^3}+\frac{3k_3^2}{k_T^4}\right.\right.\right. && \nonumber\\
    &\left.\left.\left. +\frac{8k_3^3}{k_T^5}\right)\right] -\left[k_2^2 (\bm{k_1}\cdot\bm{k_3}) \left( \frac{5-3\epsilon}{4k_T}+ \frac{5-3\epsilon}{4k_T^2} (k_1+2k_3)+ \frac{k_1k_2(5-3\epsilon)+k_3^2}{k_T^3} +\frac{3k_1k_3^2-2k_3^3}{k_T^4}-\frac{8k_3^3k_1}{k_T^5}\right)\right.\right.&& \nonumber\\
    &\left.\left. -(k_1\leftrightarrow k_2 )\right]\right]&&
\end{flalign}

\noindent where $I= \gamma_E~+~ \text{log} (-k_T\eta_{0})$ , $\eta_0 \to 0$ \footnote{\textcolor{black}{here the logarithmically divergent term comes from an integral of the type $\int_{\infty}^{\eta_0}\frac{e^{ik\eta}}{\eta}d\eta$}} and $k_T=\sum_{a=1}^3 k_a$. 
 It is easy to show that the $LR$ contribution is simply given by,
\begin{equation}
    \langle \gamma(\bm{k_1})\gamma(\bm{k_2})\zeta(\bm{k_3}) \rangle_{LR}=\frac{B_{MT}\delta P_{\zeta}(k_3)}{4P_{\zeta}(k_3)}
\end{equation}

Therefore, the final 3-point function is given by:

\begin{flalign}
  &\langle \gamma(\bm{k_1}) \gamma(\bm{k_2}) \zeta(\bm{k_3}) \rangle_{RR}= \frac{B_{MT}\delta P_{\zeta}(k_3)}{2P_{\zeta}(k_3)}- 2  \langle \gamma(\bm{k_1}) \gamma(\bm{k_2}) \zeta(\bm{k_3}) \rangle_{RR}
\end{flalign}

\textcolor{black}{where $B_{MT}$ is the mixed Bispectrum of the Maldacena cubic action which is given by \cite{Maldacena_2003}}:
\textcolor{black}{\begin{flalign}
    B_{MT}(\bm{k_1},\bm{k_2},\bm{k_3})=\frac{H^4}{M^4_pl} \frac{1}{\prod_a 2k^3_a} e^{h_1}_{ij}(\bm{k_1})e^{h_2}_{ij}(\bm{k_2}) \left(-\frac{1}{4}k^3_3+\frac{1}{2}k_3(k_1^2+k_2^2)+4\frac{k_2^2k_3^2}{k_T}\right)
\end{flalign}
}
One can easily check that:

\begin{flalign}
     \langle \gamma(\bm{k_1}) \gamma(\bm{k_2}) \zeta(\bm{k_3}) \rangle_{\bm{k_3}\to 0}= -\frac{n_t P_{\gamma}(k_1) P_{\zeta}(k_3)}{2P_{\zeta}(k_3)}- \frac{1}{2}n_t P_{\gamma}(k_1) \delta P_{\zeta}(k_3) =- n_t P_{\gamma}(k_1) \delta P_{\zeta}(k_3)
\end{flalign}
i.e the soft limit \ref{sl-1} is satisfied.\\\\

\bm{$(2)$}\bm{$\int d^{4}x\sqrt{-g} \frac{M^2_{pl}}{M^2_7} 
 {^{(3)} \!R_{ij}} {^{(3)}\!R_{ij}}$}:\\This operator gives the corrections:

\begin{flalign}
    \mathcal{O}_{\gamma \gamma \zeta}=& \int_{}{}\frac{M^2_{pl}}{M^2_7}a^{-1}\left[-\frac{1}{4} \zeta \partial^2 \gamma_{ij}\partial^2 \gamma_{ij} + \frac{\dot{\zeta}}{4H} \partial^2\gamma_{ij}\partial^2\gamma_{ij}\right] \\
     O_{\zeta\zeta}&= \int_{}{} 6 \frac{M^2_{pl}}{M^2_7} a^{-1}(\partial^2 \zeta) ^2 
    ~~~ O_{\gamma\gamma}= \int_{}{} \frac{M^2_{pl}}{4 M^2_7}a^{-1} \partial^2 \gamma_{ij}\partial^2 \gamma_{ij} 
     \end{flalign}
     The corresponding corrections to the power spectra are,
    \begin{flalign}
         \delta P_{\zeta} =15 \frac{M^2_{pl}}{M^2_7}\frac{H^4}{8M^4_{pl}\epsilon^2}~~~
    &\delta P_{\gamma} = 5 \frac{M^2_{pl}}{M^2_7} \frac{H^4}{2M^4_{pl}k^3} \delta_{h_1 h_2}
 \end{flalign}
 Again, as also noted in \cite{Bordin_2020} one should be worried about the $ \frac{\Delta\langle \zeta \zeta\rangle}{\langle \zeta \zeta \rangle_0} \sim  \frac{M^2_{pl}}{M^2_7} H^2/\epsilon M^2_{pl}$ enhancement of the power spectrum and thus we can remove the correction by taking an extra $(^{(3)}\!R)^2$ term or (for the purposes of this paper )assume that this ratio is small, which places a lower bound on $M_7$. As pointed out in \cite{Bordin_2020}, after the field redefinition $\gamma_{ij} \rightarrow \gamma_{ij} - \dot{\gamma_{ij}} \zeta/ H+...$ which removes the terms proportional to the equation of motion in the Maldacena mixed cubic action, and after integration by parts, we get:

\begin{equation}
    \mathcal{O}= \int_{}{} \sqrt{-g} \frac{M^2_{pl}}{M^2_7}{^{(3)}\!R_{ij} }{^{(3)}\!R_{ij}} +\frac{1}{2} \frac{M^2_{pl}}{M^2_7} \zeta \partial^2 \frac{\dot{\gamma_{ij} }}{H} \partial^2 \gamma_{ij} a^{-1}=-\int_{}{}\frac{1}{4}\epsilon \zeta \frac{M^2_{pl}}{M^2_7} \partial^2 \gamma_{ij} \partial^2 \gamma_{ij} a^{-1}+....\label{cabass}
\end{equation}
 After taking the soft limit $\zeta_{\bm{k_3} \rightarrow 0}$, the terms represented by dots go to 0 at leading order as mentioned in \cite{Bordin_2020}. The term proportional to $\epsilon$ however, does contribute and we have:
\begin{equation}
    \Delta_A \langle \zeta(\bm{k_3}) \gamma^{h_1}(\bm{k_ 1}) \gamma^{h_2} (\bm{k_2}) \rangle _{\bm{k_3}\to 0}= -\frac{5}{8} \frac{M^2_{pl}}{M^2_7} \left(\frac{H}{M_{pl}} \right)^6 \frac{1}{k_1^3}\frac{1}{k_3^3} \delta_{h_1 h_2}
\end{equation}

This term \ref{cabass} is not mentioned in \cite{Bordin_2020} and there, the 3-point function in the soft limit is shown to vanish at $\mathcal{O}((k_1/k_3)^0)$. The difference lies in the order of calculations, i.e. whether you compute and simplify the operator first and then do the in-in computation or just do the in-in computation for all the terms and add the answers as done in \cite{Bordin_2020}. The difference in the answers arises because $H$ is not a constant when we simplify the operators but is taken to be time-independent while doing the in-in integrals. The reason for this is that the in-in integral pick up most of the contribution near horizon crossing and therefore, it is a good approximation to take $H(t)=H_{*}$ i.e Hubble at horizon crossing.\footnote{for instance if the operator is $\mathcal{O}=\int_{}{}a^{3} M_{pl}({M_2}/{H})^2 \dot{\zeta}^3$ and we want to compute  $\langle \zeta \zeta \zeta\rangle $ we'll compute it as follows $$\langle \zeta \zeta \zeta\rangle \sim \left(\frac{M_2}{H_{*}}\right)^2 \int_{-\infty}^{0} d\eta k_1^2k_2^2 k_3^2 \eta ^4 e^{ik_T \eta}$$ i.e. we keep $H_{*}$ outside the integral. There are also multiple factors of $H_{*}$ (and $M_{pl}$), which come from $a=-1/H\eta$ and the mode functions that are also taken outside.} To get the correct soft limit, we should thus always try to eliminate $H$ dependence in the action and get everything ordered in powers of $\epsilon,\eta$, as in \ref{cabass}, as much as possible before calculating the correlator. This prescription is also followed  in standard Maldacena action calculations \cite{Maldacena_2003}. Proceeding with the calculations, we again have the ``exchange diagrams'' as before as shown below. The left one gives a contribution:

\begin{flalign}
    \Delta_B \langle \zeta(\bm{k_1}) \gamma^{h_2}(\bm{k_2}) \gamma^{h_3} (\bm{k_3}) \rangle _{\bm{k_1}\to 0}=& 2\left(\frac{B_{MT}|_{k_1 \to 0}\delta P_{\gamma}(k_1)}{2P_{\gamma}(k_1)}\right)-2\langle \zeta(\bm{k_1}) \gamma^{h_2}(\bm{k_2}) \gamma^{h_3} (\bm{k_3}) \rangle_{RR},_{\bm{k_1}\to0}\\
    &=2\left(\frac{5}{8} \frac{M^2_{pl}}{M^2_7} \frac{H^6}{M^6_{pl}} \frac{1}{k_1^3k_3^3}\delta_{h_1h_2}\right)+ \frac{15}{8} \frac{M^2_{pl}}{M^2_7} \frac{H^6}{M^6_{pl}} \frac{1}{k_1^3k_3^3}\delta_{h_1h_2} 
\end{flalign}
where the first term in the last line just follows from the Maldacena soft limit and the second term is calculated in Appendix \ref{apB}. Calculating the modified spectral tilt of $\langle \gamma \gamma \rangle$ yields:

\begin{flalign}
    \widetilde{n_t}=\frac{1}{H} \partial_t \text{log} \langle \gamma \gamma \rangle = -2\epsilon - 5\epsilon \frac{M^2_{pl}}{M^2_7} \frac{H^2}{M^2_{pl}} 
\end{flalign}
Adding both the contributions $\Delta_A$ and $\Delta_B$ gives:
\begin{flalign}
    \langle \zeta(\bm{k_1}) \gamma^{h_2}(\bm{k_2}) \gamma^{h_3} (\bm{k_3}) \rangle _{A+B, ~\bm{k_1}\to 0}= -\widetilde{n_t} (P_{\gamma}(k_1)+\delta P_{\gamma}(k_1))P_{\zeta}(k_3)
\end{flalign}
The second diagram is just like the one we considered for $^{(3)}\!R\delta K$, and hence adding it to the previous answer gives:

\begin{flalign}
    \langle \zeta(\bm{k_1}) \gamma^{h_2}(\bm{k_2}) \gamma^{h_3} (\bm{k_3}) \rangle _{\bm{k_1}\to 0}= -\widetilde{n_t} (P_{\gamma}(k_1)+\delta P_{\gamma}(k_1))(P_{\zeta}(k_3)+ \delta P_{\zeta}(k_3))
\end{flalign}    
where the equality holds at $\mathcal{O}((H/M_{pl})^6)$. These calculations for the soft limits of $\langle \gamma \gamma \zeta \rangle$ also hold for $\langle \gamma \zeta \zeta \rangle$ and one can verify it for $^{(3)}\!R \delta K$ (see Appendix \ref{apC}). Since the operator above changes $P_{\gamma}$, one might also be interested in obtaining the soft limit results for pure graviton correlators, which can be found in \cite{Cabass_2022}.

\begin{figure}[H]
\includegraphics[width=15.5cm, height=6.0cm]{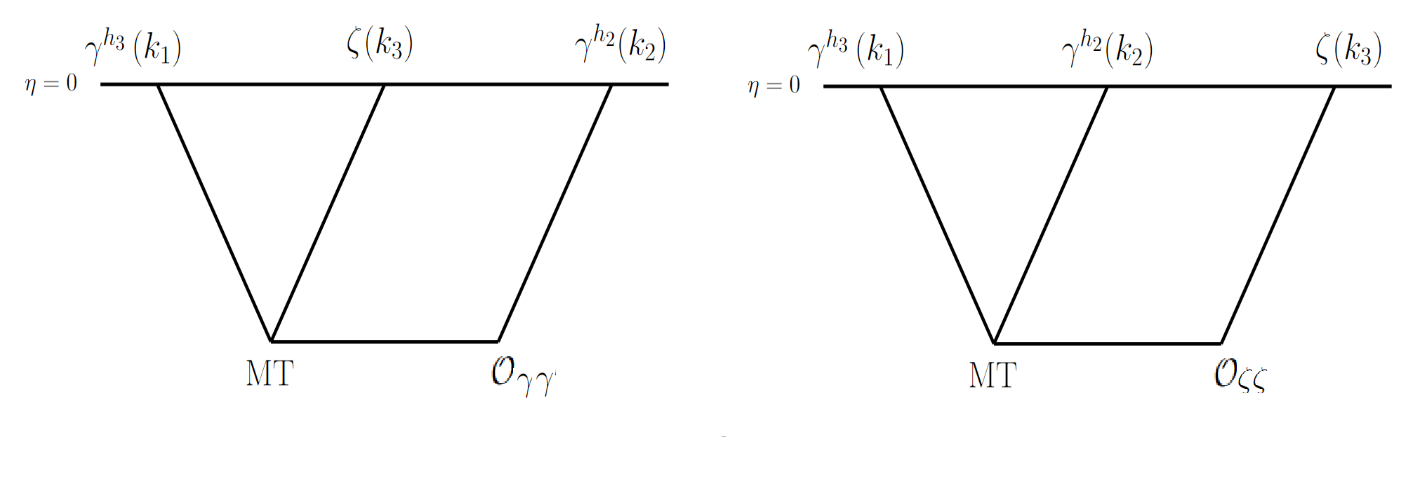}
\caption{The MT (Maldacena term) vertex is the usual cubic vertex of the Maldacena action and $\mathcal{O}_{\gamma \gamma}, \mathcal{O}_{\zeta \zeta}$ the respective quadratic correction operators from $^{(3)}\!R_{ij} ^{(3)}\!R_{ij}$ }
\end{figure}

\textbf{Purely Cubic Operators:}
We have shown how the soft limits change at the leading order in soft momenta for cubic operators. However, when we take purely cubic operators, we see from the derivative structure of $\delta K, \delta K_{ij}, ^{(3)}\!R_{ij}, ^{(3)}\!R$ (which are the building blocks for purely cubic operators) that we just have \footnote{Note that \ref{SL-2} doesn't work for operators involving terms with 4 indices like ${^{(3)}\!R_{ijkl}}$. In that case, the RHS of \ref{SL-2} is the same as \ref{SL-1}. However, in this paper, we're only dealing with operators constructed from ${^{(3)}\!R_{ij}},{^{(3)}\!R},\delta K_{ij},\delta K, \delta g^{00} $ and $ \widetilde{N}_i$ for which there's always a $\partial^2$ or $\partial_t$ acting on $\gamma_{ij}$, due to which the given limit holds. }:

\begin{flalign}
    e_3^3\langle \zeta_q \gamma \gamma \rangle|_{\bm{q} \to 0}=& e_3^3\langle\zeta_q \zeta \gamma \rangle|_{\bm{q} \to 0}=0 \label{SL-1}\\
\mathcal{B}(\gamma_q \gamma \zeta )|_{\bm{q} \to 0}=& \mathcal{B}( \gamma_q \zeta \zeta )|_{\bm{q} \to 0}=\widetilde{0}\label{SL-2}
\end{flalign}
where $\widetilde{0}$ represents the fact that the correlator is 0 as a function without taking $\bm{k_1}+ \bm{k_2} + \bm{k_3}=0$. Hence these operators only give $\mathcal{O}(q^{2})$ corrections on the RHS of the soft limits \ref{sl-1} and \ref{sl-0}. \\
\\
We have thus shown explicitly how the soft limits are obeyed for various higher derivative operators and models (i.e. $c_s=1$ or otherwise). These provide a consistency check for models beyond the Maldacena action and as we'll see next, these have an important role in the boostless bootstrap of correlators.

\section{Boostless bootstrap}
   
   Cosmological bootstrap, like any other bootstrap program, tries to fix the mathematical form of physical observables without doing explicit computations and by using very general principles such as symmetries, analytical structure, etc. As mentioned before, as of yet we have measured only the two-point scalar correlation function and its tilt \cite{Hazra_2014,Huang_2015}. The 2-point function can be bootstrapped just by using rotational, translational and scale invariance. Note that the two-point function is not exactly scale-invariant but since
   the deviation i.e the spectral tilt is very small, one can use scale invariance to fix the leading contribution (i.e. the result which survives in the pure dS limit). Since the momentum dependence is completely fixed without using dS boosts, from the bootstrap perspective it is not clear whether boosts are good symmetries of cosmological correlators or not. In single-field slow-roll models of inflation, both boost and scale invariance are broken by a slowly-evolving background (governed by $\dot{\phi}(t)$) parametrised by the slow-roll parameter $\epsilon$. This allows one to bootstrap these correlations up to slow-roll corrections i.e up to $\epsilon$ corrections using the full de Sitter isometries \cite{Pajer:2016ieg}. In terms of $\zeta$, it is not very clear that this should give the correct answer since it is a gauge mode in pure de Sitter \cite{Maldacena_2003} but one can instead bootstrap $\delta\phi$ correlators and use $\zeta=\frac{\delta\phi}{\sqrt{2\epsilon}M_{pl}}-\frac{\eta\delta\phi^2}{8\epsilon M^2_{pl}}$ to get the corresponding correlators for the curvature perturbations. In a general theory of inflation, boost breaking can be very large while still preserving approximate scale invariance. For e.g., as already mentioned in the first example of the EFToI section, $M_2$ can potentially give a speed of sound $c_s \ll 1$, thus breaking boosts by a large amount. Therefore, one can develop either a boost-preserving i.e. a conformal bootstrap \cite{https://doi.org/10.48550/arxiv.1811.00024,Shukla_2016,Rychkov_2017} or a boost-breaking or boostless bootstrap \cite{Pajer_2021,Cabass:2021fnw,https://doi.org/10.48550/arxiv.2210.02907}.\\
   The conformal bootstrap basically uses Ward Identities of Boost symmetries, along with results related to the Operator product expansion(OPE) \cite{Rychkov_2017} and soft limits to solve for the relevant correlators. The boostless bootstrap (BB), however, focuses more on the analytical structure of the correlators along with results motivated by locality, soft limits etc. A strong motivation for considering boostless theories is provided by the theorem proved in \cite{Green_2020}, which states that all non-gaussianities in $\zeta$ approximately vanish if we consider only dS invariant theories. In the upcoming sections, we explore the BB and use the rules mentioned in \cite{Pajer_2021} to construct mixed correlators from higher derivative (more than two) operators in the EFToI. 

\subsection{Relevant terms for $\langle \gamma\gamma\zeta\rangle$}
\begin{center}
\setlength{\tabcolsep}{15pt} 
\renewcommand{\arraystretch}{2.1}
\begin{longtable}{|c| c|c|c|} 
 \hline
 Term & No. of Derivatives & $\mathcal{O}(\epsilon)$ &$\mathcal{O}(\Lambda)$ or $\mathcal{O}(M_{pl})$ \\ [0.5ex] 
 \hline
 $^{(3)}\!R$ & 2 & 1 & $M_{pl}^2$ \\ 
   \hline
   $^{(3)}\!R\delta g^{00} $ & 2 & 1 & $\Lambda^2$ \\ 
   \hline
   \textcolor{blue}{$\delta K_{ij} \delta K_{ij}$} & \textcolor{blue}{3}& \textcolor{blue}{$1,\epsilon$}& \textcolor{blue}{$M^2_{pl}$} \\
    \hline
$\delta K_{ij} \delta K_{ij} \delta g^{00}$ & 3&$1$&$\Lambda^2$\\
\hline
$\delta K_{ij} \delta K_{jk} \delta K_{ki}$ & 3&$1,\epsilon$&$\Lambda$\\
\hline  
$\delta K_{ij} \delta K_{ji} \delta K$ & 3&$1,\epsilon$&$\Lambda$\\
\hline 
   \textcolor{red}{$^{(3)}\!R_{ij}\delta K_{ij}$} &\textcolor{red}{3} & \textcolor{red}{$1,\epsilon$} & \textcolor{red}{$\Lambda$} \\ 
     \hline
    $^{(3)}\!R\delta K$ & 3 & $1,\epsilon$ & $\Lambda$ \\ 
    \hline
      
     $^{(3)}\!R_{ij}\delta K_{ij}\delta g^{00}$ & 3&$1$& $\Lambda$\\
    \hline
    $^{(3)}\!R_{ij}^{(3)}\!R_{ij}  $ & 4 & 1 & 1 \\ 
     \hline
    $^{(3)}\!R_{ij}^{(3)}\!R_{ij}\delta g^{00}  $ & 4 & 1 & 1  \\ 
     \hline
    $^{(3)}\!R_{ij}\delta K_{jk}\delta K_{ki}$ & 4&$1,\epsilon$& 1 \\
    \hline
    
     $^{(3)}\!R_{ij} ^{(3)}\!R_{jk}\delta K_{ki}$ & 5&$1,\epsilon$&1/$\Lambda$ \\
    \hline
     $^{(3)}\!R_{ij} ^{(3)}\!R_{ij}\delta K$ & 5&$1,\epsilon$&$1/\Lambda$ \\
    \hline
    $^{(3)}\!R_{ij} ^{(3)}\!R\delta K_{ij}$ & 5&$1,\epsilon$&$1/\Lambda$\\
    \hline
    $^{(3)}\!R_{ij} ^{(3)}\!R_{jk}^{(3)}\!R_{ki}$ & 6&1&$1/\Lambda^2$\\
    \hline
    $^{(3)}\!R_{ij} ^{(3)}\!R_{ij}^{(3)}\!R$ & 6&1&$1/\Lambda^2$\\
   \hline
\caption{\label{tab:t-1} Quadratic and cubic operators and their contributions to $\langle \gamma \gamma \zeta \rangle $ . The 3rd column shows what powers of $\epsilon$ can be present in the terms generated by the operators, while the last column shows the coefficients of the operators in terms of some energy scale $\Lambda$. The term in red is redundant as it can be removed using $\int_{}{}A(t)^{(3)}\!R_{ij}K_{ij}=\int_{}{} A(t) ^{(3)}\!R K+  \dot{A(t)} {^{(3)}\!R}/2N + \text{total derivative}, $ while the term in blue is present in the Maldacena action. Here $\delta g^{00}=g^{00}+1$.}
\end{longtable}

\end{center}

As mentioned before, While writing the operators in Table \ref{tab:t-1}, we use covariant objects (i.e. which are covariant, at least w.r.t. the 3d metric). We also take the operators defined in \cite{Bordin_2017} :
\begin{flalign}
    V=&\frac{\dot{\delta N}-N^i\partial_i N}{N} \\
A_{\mu}=&\frac{h^{\nu}_{\mu} \nabla_{\nu}N}{N}
\end{flalign}
where $h_{\mu \nu}$ is the 3d spatial metric. Note that using $A_0$ and $V$ we can obtain $\dot{\delta N}$ and $\widetilde{N^i} \partial_i N$, so we'll use them instead. Also, using $\tilde{N_i}$ and $\partial_iN$ we can obtain $A_{i}$'s contribution to various operators. and so, we will not use $A_i$ explicitly. For $\langle \gamma \gamma \zeta\rangle$, only $\dot{\delta N}$ is relevant, which is just a higher derivative operator derived from $\delta N$ and hence it has not been included in Table \ref{tab:t-1}. All the higher derivative operators will of course be suppressed by an energy or mass scale $\Lambda$. Some of the operators give non-local terms which we shall discuss below.

\subsection{Purely Cubic Local Terms}
Let us take a local term (which is not present in the Maldacena action) as an example of bootstrapping. We shall primarily use 4 rules of bootstrapping, developed in \cite{Pajer_2021,Jazayeri_2021}. They are:

\begin{itemize}
    \item Symmetry between identical bosons. For instance, $\langle \zeta(\bm{k_1}) \gamma(\bm{k_2}) \gamma(\bm{k_3}) \rangle$ should be symmetric under $k_2 	\leftrightarrow k_3$.
    \item Amplitude rule \cite{Goodhew_2021,Maldacena_2011,Raju_2012}:
    
    \begin{equation}
\lim_{k_T \to 0} \langle \zeta_1 \zeta_2 \zeta_3..\zeta_n \rangle = \frac{(-1)^n H^{p+n-1} (p-1)! }{2^{n-1}} \frac{\text{Re}(i^{n+p+1}A_n)}{k_T^p  \prod_{a=1}^{n}k^2_a}
\end{equation}
Here $p$, for $n=3$,  is found by the total number of space and time derivatives in the interaction Hamiltonian.
\item Manifest Locality Test(MLT)\cite{Jazayeri_2021}
\begin{equation} \frac{\partial\mathcal{B}}{\partial k_1}\Bigg \vert_{k_1=0}=\frac{\partial\mathcal{B} }{\partial k_2}\Bigg \vert_{k_2=0}=\frac{\partial\mathcal{B}}{\partial k_3}\Bigg \vert_{k_3=0}=0
\end{equation}
where $\mathcal {B}(k_1,k_2,k_3) \sim ~ \langle \zeta_1 \zeta_2\zeta_3\rangle_{tr} k_1^3k_2^3k_3^3$, ``tr'' signifying that the tensor contractions have been trimmed. This equation is valid only for local operators \footnote{This can be seen easily by taking the most general local operator $\mathcal{O} \sim \zeta \cdot \zeta^{(m)} \cdot \zeta^{(n)}$ where $n,m$ denote the no. of derivatives on $\zeta$. Suppose $\bm{q}$ is the soft mode. Note that if all three $\zeta$'s had derivatives then the MLT would be trivially satisfied due to powers of $q$ coming from the derivatives. Hence the non-trivial contribution comes when we have $\bm{q}$  associated with the $\zeta$ with no derivatives. In this case, the main term to focus on is:
$$\langle \zeta_{\bm{q} \to 0} \zeta \zeta  \rangle _{tr}\sim \frac{\partial}{\partial q} \int_{-\infty}^{0}d\eta (1-iq\eta)e^{ik_T \eta }|_{\bm{q}\to 0}+ \frac{\partial}{\partial q} \int_{-\infty}^{0}d\eta (1+iq\eta)e^{-ik_T \eta }|_{\bm{q}\to 0} =0 $$
where we have omitted the mode functions involving the other 2 momenta. The analysis is easily extended to correlators with $\gamma$. }.
\item Soft limits for purely cubic operators i.e. Equations (\ref{SL-1}) and (\ref{SL-2}). We shall not bootstrap correlators arising from cubic operators since the procedure would require knowing what diagrams we need, and in this case, an explicit in-in computation might be much more simpler and economical.
\end{itemize}
We take the purely cubic vertex
\begin{equation}
    R^{(3)}_{ij} \delta K_{ij} \delta g^{00}=\int_{}{} a\frac{\Lambda}{H} \partial^2 \gamma_{ij} \dot{\gamma_{ij}} \dot{\zeta} 
\end{equation}

Direct in-in calculation gives:
\begin{equation}
    \langle \zeta(\bm{k_1}) \gamma^{h_2}(\bm{k_2}) \gamma^{h_3}(\bm{k_3}) \rangle =\frac{\Lambda}{H} e^{h_2}_{ij}(\bm{k_2})e^{h_3}_{ij}(\bm{k_3}) k^2_1 k_2^2 k_3^2 \frac{H^6}{\epsilon M^6_{pl} k^3_1 k^3_2 k^3_3}\left [ \frac{2}{(k_1+k_2+k_3)^3} + 3\frac{k_2+k_3}{(k_1+k_2+k_3)^4}\right]  \label{e-1}
\end{equation} 

we have $p=4$ for this vertex and \footnote{note that there's no way to bootstrap the amplitude since any function of the form $[23]^4 f(k_1,k_2,k_3)$, where f is a degree 4 polynomial, is a valid $p=4$ amplitude. Here [ ] is the relevant helicity bracket.}
\begin{equation}
    A[1^{0},2^{h_2},3^{h_3}] \sim e^{h_2}_{ij}(\bm{k_2}) e^{h_3}_{ij}(\bm{k_3})k_1 k_2 k_3 (k_2+k_3)
\end{equation}
where we have kept the normalization arbitrary (which contains information about things like $\Lambda$ and $\epsilon$ dependence). From this we write an ansatz using the first rule(where $k_T=k_1+k_2+k_3$ , $e_1=k_2+k_3$, $e_2=k_2k_3$ and $e_3=k_1k_2k_3$):
\begin{equation}
\begin{split}
    \langle \zeta(\bm{k_1}) \gamma&(\bm{k_2}) \gamma(\bm{k_3}) \rangle \sim \frac{e^{h_2}_{ij}(\bm{k_2})e^{h_3}_{ij}(\bm{k_3})}{k^4_Te^3_3} \left [ e_3 k_1 k_2k_3(k_2+k_3) + k_T (A_1e_2^3 + A_2 e_2^2e_1^2+ A_3 e_2 e_1^4+A_4 e_1^6) +\right.\\
    &~~\left. k_T^2 (A_5 e_2^2e_1+ A_6e_2e_1^3+A_7e_1^5)+ k_T^3 (A_8e_2^2+ A_9e_2e_1^2+A_{10} e_1^4)+ k_T^4 (A_{11}e_2 e_1 +A_{12}e_1^3)\right.\\
    &\left.+ k_T^5 (A_{13}e_2+ A_{14} e_1^2 )+ A_{15} k_T^6 e_1 + A_{16} k_T^7 \right]\\
 \end{split}
\end{equation}

We get the following set of equations after applying MLT and soft limits for various momenta:
\begin{flalign}
    &\text{Soft limit for $k_2,k_3$:} ~~~A_4=A_7=A_{10}=A_{12}=A_{14}=A_{15}=A_{16}=0\\
    &\text{MLT for $k_2,k_3$:} 
    ~~~A_3=A_6=A_9=A_{11}=A_{13}=0\\
    &\text{Soft limit for $k_1$:} ~~~A_2+A_5+A_8=0\\
    &\text{MLT for $k_1$:} 
    ~~~A_1=0 ~~ 3A_2+2A_5+A_8=0 
\end{flalign}
which fixes our correlator to be:
\begin{equation}
     \langle \zeta(\bm{k_1}) \gamma(\bm{k_2}) \gamma(\bm{k_3}) \rangle \sim \frac{e^{h_2}_{ij}(\bm{k_2})e^{h_3}_{ij}(\bm{k_3})}{k^4_Te^3_3} e_3 ^2 \left [ A_2 k_T + e_1\right]
\end{equation}
which means we're able to fix the bispectra up to an overall factor and another arbitrary constant. This expression agrees with the explicit calculation \ref{e-1} with $A_2=2/3$.

\subsection{Non-local terms}

We have the following non-local terms (see Appendix \ref{apD}) at various orders in $\epsilon$ and energy scales:
\begin{center}
\begin{tabular}{|c|c|c|}
    \hline
    Operator & $\mathcal{O}(\epsilon)$ & $\mathcal{O}(H/M_{pl})$ or $\mathcal{O} (H/\Lambda)$ \\
    \hline
    
     $\delta K_{ij}\delta K_{ij}$  & $\epsilon$  & $H/M_{pl}$ \\
     $\delta K_{ij}\delta K_{jk} \delta K_{ki}$ & $\epsilon$& $(H/M_{pl})^2 (\Lambda/ M_{pl})$\\
     $\delta K_{ij}\delta K_{jk} {^{(3)}\!R}_{ki}$& $\epsilon$ & $(H/M_{pl})^3$\\
     $\delta K_{ij} {^{(3)}\!R}_{jk}{^{(3)}\!R}_{ki}$& $\epsilon$ & $(H/M_{pl})^3 (H/\Lambda)$\\
     \hline
\end{tabular}
\end{center} 

(For the sake of brevity we have not included the odd parity terms but their contributions are of the same order as the last 3 terms). The second term in the table gives:
\begin{equation}
    \int_{} \sqrt{-g} ~ \Lambda\delta K_{ij}\delta K_{jk} \delta K_{ki}= \int_{} -\frac{3}{4} 
a^{3} \Lambda \epsilon \dot{\gamma_{ij}}\dot{\gamma_{jk}} \partial_i\partial_k \partial^{-2}\dot{\zeta}\label{nlocal}
\end{equation} 
\\[0.3cm]
for which the explicit in-in calculation yields:
\begin{equation}
    \langle \zeta(\bm{k_1}) \gamma^{h_2}(\bm{k_2}) \gamma^{h_3}(\bm{k_3})\rangle= -\frac{3}{2} \frac{H^6}{M_{pl}^6}\left(\frac{\Lambda}{H}\right) e^{h_2}_{ij}(\bm{k_2})e^{h_3}_{jm}(\bm{k_3}) k_{1i}k_{1m} \frac{k_2^2k_3^2}{ k_T^3 e_3^3}\label{e-2}
\end{equation}

The non-local term in \ref{nlocal} naively doesn't seem to have a proper flat space counterpart, but as pointed out in \cite{Pajer_2021}, it can be considered to come from a toy model:\\[0.3cm]
\begin{equation}
S_{flat}=\int_{}{} d^4x ~~\epsilon (\partial_{\mu} \zeta) ^2 - \frac{1}{2} (\partial_i X)^2 + \frac{M_{pl}^2}{8} (\partial_{\mu} \gamma_{ij}) ^2 - \frac{3\Lambda \epsilon \dot{\zeta_0}}{8} X + \frac{3 \Lambda \epsilon}{8 \dot{\zeta_0}}X\dot{\zeta}^2 +\Lambda_{}\dot{\gamma_{ij}}\dot{\gamma_{jk}} \partial_i\partial_k X
\end{equation}

Here $\zeta_0(t)=-(\sqrt{2\epsilon})^{-1}\bar{\phi}(t)$ is the background value of the scalar field. Integrating out the field $X$ above gives us an EFT with the desired non-local term, which gives an amplitude (with no of derivatives, $p=3$):

\begin{equation}
A[1^0,2^{h_2},3^{h_3}]= e^{h_2}_{ij}(\bm{k_2})e^{h_3}_{jm}(\bm{k_3}) k_{1i}k_{1m}\frac{k_2k_3}{k_1} 
\end{equation}
Taking the ansatz as before(the soft limit for $k_2,k_3$ has already been taken):
\begin{equation}
\langle \zeta(\bm{k_1}) \gamma(\bm{k_2}) \gamma(\bm{k_3}) \rangle \sim \frac{e^{h_2}_{ij}(\bm{k_2})e^{h_3}_{jm}(\bm{k_3}) k_{1i}k_{1m}}{k_T^3 e_3^3} \left [ \frac{k_2k_3}{k_1}e_3 + Ak_Te_1e_2+ Bk_T^2e_2 \right]
\end{equation}
applying MLT with respect to $k_2$ and $k_3$ fixes $A=B=0$ and hence, the correlator up to an overall factor, matching with \ref{e-2}. Note that as the value of $p$ increases, we'll get more and more unknown parameters. Specifically for the non-local terms, we cannot use the MLT w.r.t the non-local momenta, which removes one condition and increases the no. of arbitrary constants. \textcolor{black}{Define $P=p-m$, where $m$ is the number of external momenta present in the tensor contractions. It is easy to see that $P$ can only take the values $p,p-2$ and $p-4$ (in the bootstrap example above, we had $P=p-2$). From the conditions we have used, one can simply find that:}

\begin{itemize}
    \item \textcolor{black}{For local terms and odd $p$ with $p \geq 3$, one has at least $(p-3)(p+1)/4 $ parameters if $P=p$ and at least $(P-1)^2/4$ parameters otherwise, that can't be fixed (excluding the overall factor). For even $p$, this number is $(p^2-2p-4)/4$ for $P=p$ and $P(P-2)/4$ otherwise. }
    
    \item  \textcolor{black}{For non-local terms and odd $p$ with $p \geq 3$, one has at least $(p^2-5)/4 $ parameters if $P=p$ and at lea st $(P^2-1)/4$ parameters otherwise, that can't be fixed (excluding the overall factor). For even $p$, this number is $(p^2-4)/4$ for $P=p$ and $P^2/4$ otherwise}.
\end{itemize}
We want to point out that the source of these non-localities is rooted in the fact that not all metric components are dynamical variables. Since, we have constraint equations for these non-dynamical variables, one plugs in their formal solution in the action which can potentially involve inverse differential operators since the constraint equations are differential equations. 

\subsection{Extending results to $\langle \gamma \zeta \zeta\rangle$}
All operators in this case have enough derivatives on $\gamma$ and $\zeta$ so that the soft limits \ref{SL-1}, \ref{SL-2} are still valid. A similar bootstrap analysis can be carried out for non-local and local terms separately. One again finds that the non-local terms start appearing at $\mathcal{O(\epsilon)}$. The operators are summarized in  Table \ref{table2} below
\\[0.1cm]

\begin{center}
\setlength{\tabcolsep}{15pt} 
\renewcommand{\arraystretch}{2.2}
\begin{longtable}{|c| c|c|c|} 
 \hline
 Term & No. of Derivatives & $\mathcal{O}(\epsilon)$ &$\mathcal{O}(\Lambda)$ or $\mathcal{O}(M_{pl})$ \\ [0.5ex] 
 \hline
 \textcolor{blue}{$c(t)\delta g^{00} $} & \textcolor{blue}{0} &\textcolor{blue} {$\epsilon$} & \textcolor{blue}{$M^2_{pl}$} \\ 
   \hline
\textcolor{blue}{$\delta K_{ij} \delta K_{ij}$} & \textcolor{blue}{2}& \textcolor{blue}{$1,\epsilon$}& \textcolor{blue}{$M^2_{pl}$} \\
    \hline
    
   $\delta K_{ij}\widetilde{N_i}\partial_j\delta g^{00}$ & 2&$1,\epsilon$& $\Lambda^2$ \\
    \hline
    
   \textcolor{red}{$^{(3)}\!R_{ij}\delta K_{ij}$} &\textcolor{red}{3} & \textcolor{red}{$1,\epsilon$} & \textcolor{red}{$\Lambda$} \\ 
     \hline
$\delta K_{ij} \delta K_{jk} \delta K_{ki}$ & 3&$1,\epsilon,\epsilon^2$&$\Lambda$\\
\hline  
$\delta K_{ij} \delta K_{ji} \delta K$ & 3&$1,\epsilon,\epsilon^2$&$\Lambda$\\
\hline     $^{(3)}\!R_{ij}\delta K_{ij} \delta g^{00}$ & 3 & $1,\epsilon$ & $\Lambda$ \\ 
    \hline

      $\delta K_{ij}\partial_i \delta g^{00}\partial_j\delta g^{00}$ & 3&1& $\Lambda$ \\
    \hline$^{(3)}\!R_{ij}\widetilde{N_i}\partial_j\delta g^{00}$ & 3&$1,\epsilon$& $\Lambda$ \\
    \hline
    
    $^{(3)}\!R_{ij}\delta K_{ij}\delta K  $ & 4 & $1,\epsilon,\epsilon^2$ & 1  \\ 
     \hline
    $^{(3)}\!R_{ij}\delta K_{jk}\delta K_{ki}$ & 4&$1,\epsilon,\epsilon^2$& 1 \\
    \hline
    
     $^{(3)}\!R_{ij} \partial_i\delta g^{00} \partial_j g^{00}$ & 4&1&1 \\
    
    \hline
    $^{(3)}\!R_{ij} ^{(3)}\!R_{ij}\delta g^{00}$ & 4&1&$1/\Lambda$\\
    \hline
    $^{(3)}\!R_{ij} ^{(3)}\!R_{jk}\delta K_{ki}$ & 5&$1,\epsilon$&$1/\Lambda$\\
    \hline
    $^{(3)}\!R_{ij} ^{(3)}\!R_{ij}\delta K$ & 5&$1,\epsilon$&$1/\Lambda$\\
    \hline 
    $^{(3)}\!R_{ij} ^{(3)}\!R_{jk}^{(3)}\!R_{ki}$ & 6&1&$1/\Lambda^2$\\
    \hline
    $^{(3)}\!R_{ij} ^{(3)}\!R_{ij}^{(3)}\!R$ & 6&1&$1/\Lambda^2$\\
    \hline 
\caption{Quadratic and cubic operators and their contributions to $\langle \gamma \zeta \zeta \rangle $. Again, the blue terms are Maldacena terms, and the red term is removable by the identity mentioned below Table \ref{tab:t-1}.} \label{table2}
\end{longtable}
\end{center}

We take the term (which is the operator $\mathcal{O}_3$ in Appendix \ref{apD}) :
\begin{equation}
    \mathcal{O} = \delta K_{ij}\partial_i \delta g^{00}\partial_j\delta g^{00} \sim\int_{} a \frac{1}{\Lambda} \dot{\gamma_{ij}} \partial_i \dot{\zeta} \partial_j \dot{\zeta}     
\end{equation}
The explicit in-in calculation gives:
\begin{equation}
\langle \gamma^{h}(\bm{k_1}) \zeta(\bm{k_2}) \zeta(\bm{k_3})\rangle= 6 \frac{H^7}{\epsilon^2 M_{pl}^6 \Lambda} \frac{e_{ij}^{h}(\bm{k_1})k_{2i}k_{3j}}{e_3 k_T^5} \label{expl}
\end{equation}
We have the corresponding flat space amplitude:
\begin{equation}
    A[1^{h},2^{0},3^{0}] \sim e_3 e^h_{ij} (\bm{k_1})k_{2i}k_{3j}
\end{equation}
which gives us the ansatz for the correlator to be:

\begin{flalign}
    \langle \gamma^{h}(\bm{k_1}) \zeta(\bm{k_2}) \zeta(\bm{k_3})\rangle\sim  \frac{e^h_{ij}(\bm{k_1}) k_{2i}k_{3j}}{k_T^5 e_3^3} \left[e_3^2 + k_T (A_1 e_2^2e_1 +A_2 e_2e_1^3 +A_3 e_1^5) + k_T^2 (A_4 e_2^2 + A_5 e_2 e_1^2 + A_6 e_1^4) \right. \nonumber\\
    \left. k_T^3 (A_7 e_2 e_1 +A_8 e_1^3)+ k_T^4 (A_9 e_2+ A_{10} e_1^2)+ A_{11} k_T^5 e_1 +A_{12}k_T^6
    \right]
\end{flalign}
Soft limits and MLTs give the following equations:

\begin{flalign}
    \text{Soft limit $\bm{k_1}\to 0$} &\begin{cases}
    A_1+A_4=0 \\
    A_2+A_5+A_7+A_9=0\\
    A_3+A_6+A_8+A_{10}+A_{11}+A_{12}=0
    \end{cases}
\end{flalign}

\begin{flalign}
    \text{MLT for $k_1$} &\begin{cases}
        4A_1+3A_4=0\\
        4A_2+3A_5+2A_7+A_9=0 \\
        4A_3+ 3A_6+ 2A_8+A_{10}-A_{12}=0
    \end{cases}
    \\
    \text{Soft limit $\bm{k_2}\to 0$} &
    ~~\text{Already satisfied due to the tensor structure}\\
    \text{MLT for $k_2$} &\begin{cases}
        5A_3+A_2+2A_6=0\\
        A_3=0 \\
        A_5+4A_6+ 3A_8=0\\
        A_7+3A_8+4A_{10}=0\\
        A_9+2A_{10}+5A_{11}=0\\
        A_{11}+6A_{12}=0
    \end{cases}
\end{flalign}
which leads to the following correlator with just one arbitrary parameter:
\textcolor{black}{\begin{equation}
\begin{split}
    \langle \gamma^{h}(\bm{k_1}) \zeta(\bm{k_2}) \zeta(\bm{k_3})\rangle\sim  \frac{e^h_{ij}(\bm{k_1}) k_{2i}k_{3j}}{k_T^5 e_3^3}&\left[ e_3^2+ A_{10} \left( -2 k_T e_2 e_1^3 +k_T^2( 2e_2e^2_1+ e_1^4)+k_T^3(e_2e_1-2e_1^3) \right.\right.\\
    &\left.\left.+ k_T^4(-2e_2+e_1^2) \right)\right]
\end{split}
\end{equation}}
which matches the explicit result \ref{expl} for $A_{10}=0$.
\section{Going to Bogolyubov states}
In our calculations, we are not compelled to fix the initial condition (and the subsequent evolution) by taking the Bunch-Davies (BD) vacuum. \textcolor{black}{A simple family of excited states one might be motivated to take is the family of states that are Bogolyubov transformations of the BD vacuum, since they just involve taking a linear combination of the creation and annhilation operators of BD. In particular, within this family, one can take the well-known family of $\alpha$ vacua }\cite{Allen:1985ux,Shukla_2016} since they respect the symmetries of the quasi dS background. Bootstrapping $\alpha$-vacua answers directly using the BB is difficult since we don't have the soft limit conditions such as \ref{SL-2}, because of the pole structures mentioned in Section \ref{sec-3}. However, once we have bootstrapped $\mathcal{B}(k_1,k_2,k_3)$(as defined above) for BD, we can extend the result to $\alpha$ vacua easily. For a (k-independent) Bogolyubov transformation (BT), just by noting the form of the mode functions, which are given by :

\begin{equation}
\begin{split}
    u_k(\eta)=\alpha &(1-ik\eta)e^{ik\eta}+\beta(1+ik\eta)e^{-ik\eta} \\
    &\abs{\alpha}^2-\abs{\beta}^2=1
\end{split} 
\end{equation}

\textcolor{black}{we can give an ansatz for the BT bispectra as follows}:

\textcolor{black}{\begin{equation}
\begin{split}
\mathcal{{B}}_{BT} (k_1,k_2,k_3,\{\bm{k}\})= \text{Re}&\left[ \left(\alpha+\beta \right)^3\left( \psi_3'(k_1,k_2,k_3,\{\bm{k}\}) {\alpha}^{*3}+ \sum_{\text{cyclic}}\psi_3' (-k_1,k_2,k_3,\{\bm{k}\}) {\alpha}^{*2} \beta^* \right. \right.\\
&\left.\left.+ \sum_{\text{cyclic}}\psi_3' (-k_1,-k_2,k_3,\{\bm{k}\}) \alpha^{*} \beta^{*2}+\psi_3' (-k_1,-k_2,-k_3,\{\bm{k}\})\beta^{*3}\right)\right] \label{an}
\end{split}
\end{equation}}

\textcolor{black}{where $\psi_3'$ is the trimmed cubic wavefunction coefficient in BD vacuum \cite{Goodhew_2021,Cabass:2021fnw}. From the cosmological optical theorem, if we have odd parity interactions i.e. odd number of momenta contracted with the polarization tensors, the correlator for BD is 0 and we need the wavefunction coefficients to get the final answer for BT states. However, for even parity interactions, we have $\mathcal{B}_{BD}=\psi_3'$ and in this case we can get the answers for BT states directly from the BD answers.} Putting $\alpha=\text{cosh} \alpha$ and $\beta=  i ~\text{sinh}\alpha$ , we get the $\alpha$ vacua result. 
Using this equation  to bootstrap $\langle \gamma \gamma \gamma\rangle$ in $\alpha$ vacua for the Maldacena action, we take the well-known result for BD which was bootstrapped in \cite{Pajer_2021} (also explicitly calculated in \cite{Maldacena_2003}), and get :

\begin{equation}
\begin{split}
  \langle \gamma ^{h_1}(\bm{k_1})&\gamma ^{h_2} (\bm{k_2})\gamma ^{h_3} (\bm{k_3})\rangle_{\alpha}=-\frac{2H^4}{M_{pl}^4 }\frac{1}{(\prod^{3}_{a=1} k_a^3)}e^{h_1}_{ii'}e^{h_2}_{jj'}e^{h_3}_{kk'}t_{ijk}t_{i'j'k'} \left [\left(-k_T+ \frac{\sum k_i k_j}{k_T} +\frac{k_1k_2k_3}{k_T^2}\right) + \right. \\
  &\left. \text{sinh}^2 2\alpha \left( -(-k_1+k_2+k_3) + \frac{k_2k_3-k_1k_2-k_1k_3}{(-k_1+k_2+k_3)} -\frac{k_1k_2k_3}{(-k_1+k_2+k_3)^2}\right)+\text{cyclic}\right] \\\\
  &~~~~~~~~~~~~~~~~~~~~t_{ijk}= k_{2i}\delta_{jk}+k_{3j}\delta_{ki}+k_{1k}\delta_{ij}
\end{split}
\end{equation}
which agrees with the explicit in-in result in \cite{Maldacena_2011,Kanno_2022}. We also get the following expression for the pure scalar correlator:

\begin{equation}
  \begin{split}
&\langle\zeta(\bm{k_1}) \zeta(\bm{k_2})\zeta(\bm{k_3})\rangle_{\alpha}=\frac{H^4}{32 M_{pl}^4 \epsilon^2}\frac{1}{ (\prod_{a=1}^{3}k_a^3)} \left[2(\epsilon-\eta)\sum_{a} k^3_a+ \epsilon \left (\sum_{a} k^3_a+ \sum_{a \neq b} k^2_a k_b+ 8\sum_{a>b}\frac{k_a^2 k_b^2}{k_T}\right)+\right.\\
&\left. \text{sinh}^2 2\alpha \left( 2(\epsilon-\eta)\sum_a k_a^3+\epsilon \left(\sum_{a}k_a^3+ \sum_{a \neq b} k^2_a k_b+ \sum_{a>b} 8\frac{k_a^2 k_b^2}{k_2+k_3-k_1} + 8\frac{k_a^2 k_b^2}{k_3+k_1-k_2} + 8\frac{k_a^2 k_b^2}{k_1+k_2-k_3} \right)\right)\right]  
  \end{split}
\end{equation}

which agrees with the calculation done in \cite{Shukla_2016}. This demonstrates that the precription given above indeed works.

Similarly, using the BD results, we get the following for mixed correlators for Maldacena action in $\alpha$ vacua:

\begin{equation}
   \begin{split} \langle\gamma^h(\bm{k_1})&\zeta(\bm{k_2})\zeta(\bm{k_3})\rangle_{\alpha}= \frac{H^4}{4M^4_{pl} \epsilon} \frac{1}{(\prod_{a=1}^{3}k_a^3)} e^h_{ij}(k_1) k_{2i}k_{3j} \left[\left(-k_T+ \frac{\sum k_i k_j}{k_T} +\frac{k_1k_2k_3}{k_T^2}\right) + \right.\\
  &\left. \text{sinh}^2 2\alpha \left( -(-k_1+k_2+k_3) + \frac{k_2k_3-k_1k_2-k_1k_3}{(-k_1+k_2+k_3)} -\frac{k_1k_2k_3}{(-k_1+k_2+k_3)^2}\right)+\text{cyclic}\right] 
   \end{split}
\end{equation}

\begin{equation}
    \begin{split}
\langle&\zeta (\bm{k_1})\gamma^{h_2}(\bm{k_2})\gamma^{h_3}(\bm{k_3})\rangle_{\alpha}= \frac{H^4}{8M^4_{pl}} \frac{1}{(\prod_{a=1}^{3}k_a^3)} e^{h_2}_{ij}(k_2) e^{h_3}_{ij}(k_3) \left[ \left(-\frac{1}{4}k_1^3+ \frac{1}{2}k_1 (k_2^2+k_3^2) + 4\frac{k^2_2k^2_3}{k_T}\right)\right. \\
&\left. + \text{sinh}^2 2\alpha \left( -\frac{1}{4}k_1^3+ \frac{1}{2}k_1 (k_2^2+k_3^2) + 4\frac{k^2_2k^2_3}{(k_2+k_3-k_1)}+ 4\frac{k^2_2k^2_3}{(k_3+k_1-k_2)}+ 4\frac{k^2_2k^2_3}{(k_1+k_2-k_3)}\right) \right]
    \end{split}
\end{equation}

\textcolor{black}{To take an example of an odd parity interaction we can take the Weyl action and calculate the un-trimmed wavefunction coefficient $\psi_3$ (up to some numerical factors): }

\textcolor{black}{\begin{align}
    S_3=&\int_{}{} d\eta d^3 x a^{-5} \left(\partial_{\eta}\Pi^{+}_{ij} \partial_{\eta}\Pi^{+}_{jk}\partial_{\eta}\Pi^{+}_{ki} - \partial_{\eta}\Pi^{-}_{ij} \partial_{\eta}\Pi^{-}_{jk}\partial_{\eta}\Pi^{-}_{ki} \right)\\
    &\text{where}~~\partial_\eta\Pi^{\pm}_{ij}=\frac{1}{2}\left(\partial_{\eta}(a\partial_\eta \gamma_{ij}) \mp i\epsilon_{jab}\partial_{b}\partial_{\eta}\gamma_{ia}  \right)\\
    &\psi_3 \sim \frac{k_1^2k_2^2k_3^2}{k_T^6} \left(\epsilon_{iab}\epsilon_{jcd}\epsilon_{kfg}e^{h_1}_{jb}e^{h_2}_{kd}e^{h_3}_{ig} k_{1a}k_{2c}k_{3f}-\sum_{cyclic} k_2k_3 \epsilon_{jab}k_{1b}e^{h_1}_{ia}e^{h_2}_{ki}e^{h_3}_{kj}\right)\label{tri}
\end{align}}

\textcolor{black}{We can simplify the last equation using the relation: $\epsilon_{iab}k_{b}e^{h}_{ja}=-ikhe_{ij}$. However, we must keep in mind that while using the ansatz \ref{an} we have to take the trimmed part as the one before we use the relation above to simplify \ref{tri}, i.e we consider the unsimiplified levi-cevita contractions to be the tensor contractions. Hence we don't flip the signs of $k$'s generated from this contraction while using the ansatz. We also note that \ref{tri} has multiple tensor contractions and for each contraction, the trimmed part obeys \ref{an}, so we can just add up the answers. This finally gives the following correlators for arbitrary helicities $h_i=\pm 1$}:

\textcolor{black}{\begin{align}
\langle \gamma^{h_1} (\bm{k_1}) \gamma^{h_2} (\bm{k_2})\gamma^{h_3} (\bm{k_3})\rangle_{\alpha} \sim ~&\text{sinh}4\alpha e^{h_1}_{ij}e^{h_2}_{jk}e^{h_3}_{ki}\left[\frac{3}{k_T^6} (h_1+h_2+h_3+h_1h_2h_3)\right. \nonumber \\ 
&\left.-\sum_{cyclic}\frac{1}{(-k_1+k_2+k_3)^6} (h_1-h_2-h_3+h_1h_2h_3) \right]\\
\implies\langle \gamma^{-} (\bm{k_1}) \gamma^{+} (\bm{k_2})\gamma^{+} (\bm{k_3})&\rangle_{\alpha}  \sim  \sinh4\alpha e^{h_1}_{ij}e^{h_2}_{jk}e^{h_3}_{ki} \frac{1}{(k_2+k_3-k_1)^6}=-\langle\gamma^{+} (\bm{k_1}) \gamma^{-} (\bm{k_2})\gamma^{-} (\bm{k_3})\rangle_{\alpha}
\end{align}}
\textcolor{black}{Again, these results match with the explicit calculations done in \cite{gong2023new}.} \textcolor{black}{While all the calculations above were done for cases where the Bogolyubov coefficients $\alpha,\beta$ are $k$-independent, one can easily generalise \ref{an} to cases where they are $k$-dependent. In that case, one would have to flip the signs of $k$'s for $\alpha,\beta$ like we did for $\psi$ in \ref{an}.}\\

We see that the prescription saves us the effort of doing the cumbersome in-in calculation which involves simplifying a lot of integrals. 
Note that, our prescription is for {\it the inflationary correlations
functions} and works for general Bogolyubov states (thus, goes beyond the
pure de-Sitter/CFT results in $\alpha$ vacua \cite{Jain:2022uja}).


\section {Conclusions}
In this article, we aimed to understand the mixed graviton and scalar bispectra in the EFT of inflation. A summary of the main results of the paper is as follows:

\begin{itemize}
    \item Following the methods prescribed in  \cite{Cheung_2008}, we wrote a general EFToI and attempted to organize terms in the order of the number of derivatives on them w.r.t the metric perturbations. We also clarified the energy scale counting in terms of $H/M_i$ where $M_i$'s are the UV cutoffs/high mass scales appearing in our EFTs. 

    \item We gave some general constraints on the EFT parameters, namely the bounds due to small spectral tilt, unitarity bound in flat space limit and experimental bounds of non-gaussianities \cite{Cheung_2008}. We gave 2 simple examples where these bounds constrained some of the arbitrary EFT coefficients. 
    
    \item We explicitly checked the soft limits \ref{sl-2}, \ref{sl-1} and \ref{sl-0} for EFT operators, which change both the quadratic and cubic action for $\zeta$ or $\gamma$. These limits as we checked, are obeyed for cubic operators at leading order in soft momenta  and leading order in the ``couplings'' (i.e. $\epsilon, \eta$ etc.) and $H/M_{pl}$. We clarified some confusion in the literature related to what diagrams to take and how to organise terms in order to get the correct soft limits. Hence, as expected from the general derivation of the soft limits \cite{Maldacena_2003,Creminelli_2012} (also see Section \ref{sec-3}), they can be extended to models beyond the standard Maldacena action \cite{Maldacena_2003}.
    
    \item We attempted to bootstrap the three-point correlators from purely cubic operators, i.e. operators that do not change the quadratic action, by noticing that they don't contribute to the soft limit at leading order in soft momenta (see Equations \ref{SL-1}, \ref{SL-2}). Using the bootstrap prescription in \cite{Pajer_2021}, we found (as expected) that the number of undetermined parameters increases with the number of derivatives. Furthermore, this bootstrap method heavily relies on knowing the interaction hamiltonian since we use the amplitude of the interaction to determine the residue of the total energy pole \cite{Goodhew_2021}. Hence, the BB method is more of an alternative to doing the in-in integrals than an ideal ``boundary perspective'' bootstrap.
    \item Since we are not compelled to fix BD initial conditions, we would be interested in extending BD results to more arbitrary vacua. We give an ansatz through which the results of BD can be extended to give the 3-point correlators for BT states in cases where the interaction hamiltonian has even parity. For odd parity cases, we still have a relation in terms of the wavefunction coefficients. This helps us in bypassing a lot of in-in integrals. In particular, we derive the 3-point correlators for $\alpha$ vacua (a subset of BT states) to show how useful the ansatz is.
    \end{itemize}
It will be interesting to explore mixed quartic operators for and beyond the Maldacena action and check the soft limits for these since some exchange diagrams also come into the picture here as they're of a similar order in ``couplings'' and $ H/M_{pl}$ as contact diagrams. The right-hand side of the soft limits might be tricky to evaluate due to the momentum dependence of the polarization tensors. We would also like to extend our $\alpha$ vacua results to four-point correlators and explore the implications of the new kinds of pole structures we get.

\section*{Acknowledgments}
DG acknowledges support through the Ramanujan Fellowship
and MATRICS Grant of the Department of Science and Technology, Government of India. We thank Enrico Pajer
for clarifications related to soft limits via email and for useful comments on the first arXiv version of our work.
We also thank Sachin Jain and Muhammad Ali for useful discussions. 

\begin{appendices}

\section{Calculating the action in Unitary Gauge}\label{apA}

From the definitions of the ADM metric variables, we have  $\widetilde{N^i}=-g_{00}N^i= N^2 N^i$. We further write $N=1+ \delta N$ and then consider the following quadratic action:
\begin{flalign}
    \mathcal{L}_2&=\sqrt{-g }M^2_{pl}\left ( \frac{1}{2}R^{(4)}+ m_1\delta K^i_j \delta K^j_i + m_2(\delta K)^2 + D\delta K -M^2_{pl}\lambda(t) -c(t)g^{00} \right.  &&\nonumber\\
    &\left. +M_1 g^{ij} \partial_i (g^{00}+1)\partial_j (g^{00}+1) +
     M^2_2 (g^{00}+1)^2 + M_3 (g^{00}+1)\delta K+ m_3 ^{(3)}\!R ( g^{00}+1)\right)
\end{flalign}
where $c(t)$,$\lambda(t)$  are as defined before. Taking $\widetilde{N^i}=\partial_i \chi$, the equations of motion for $N$ and $\widetilde{N^i}$ gives 

\begin{flalign}
\delta N &\left((m_1+ 3m_2-1)H-M_3\right) + (m_1+m_2)\partial^2 \chi= \dot{\zeta}(m_1+3m_2-1) &&\\ 
\partial^2 \chi &=\frac{-1}{(m_1+ 3m_2-1)H-M_3} \left( -\partial^2 \zeta a^{-2}+ c(t)\delta N + 4M^2_2 \delta N -4a^{-2} M_1 \partial^2 \delta N \right) -3\dot{\zeta}&&\\
&+\frac{3\dot{\zeta}H(m_1+3m_2-1)}{(m_1+3m_2-1)H-M_3} -\frac{3\dot{\zeta}H(m_1+3m_2-1)M_3}{((m_1+3m_2-1)H-M_3)^2}&&\nonumber
\end{flalign}
To solve these equations, one will have to take $m_1+m_2=0$ so that the equations separate.

\section{Calculating the RR vertex for $^{(3)}\!R_{ij}^{(3)}\!R_{ij}$}\label{apB}

The expression for the diagram where both the vertices are time ordered, i.e. RR vertices \cite{Weinberg_2005} after taking the soft limit is given by:
\begin{flalign}
\langle &\gamma^{h_1}(\bm{k_1} ) \gamma^{h_2}(\bm{k_2} ) \zeta(\bm{k_3})|_{\bm{k_3} \to 0}\rangle |_{RR}= \frac{H^6}{M^6_{pl}k_1^9 4\epsilon k_3^3} \left( \frac{M_{pl}^2 \epsilon}{M_7^2 (4\cdot 8)} 2\cdot 2\cdot 2 \cdot 2 ~ e^{h_1}_{ij}(\bm{k_1}) e^{h_2}_{ij}(\bm{k_2}) \right) \nonumber\\
&\left[ \int_{-\infty}^{0}\left(k_1^2k_2^2 e^{2ik_1\eta}-( k_1 \cdot k_2) \frac{(1-ik_1\eta)(1-ik_2\eta)}{\eta^2} e^{2ik_1\eta} \right) \left(\int_{\eta}^{0}(1+ik_2\eta') k_2^4 (1-ik_2\eta')\right) d\eta d\eta'+\right.\nonumber\\
&\left.
\int_{-\infty}^{0}\left(k_1^2k_2^2 -( k_1 \cdot k_2) \frac{(1-ik_1\eta)(1+ik_2\eta)}{\eta^2}  \right) \left(\int_{-\infty}^{\eta}(1-ik_2\eta')^2 k_2^4 e^{2ik_2\eta}\right) d\eta d\eta'
\right]
\end{flalign}

where we have taken $\bm{k_3}\to 0$ but have not yet put $\bm{k_1}=\bm{-k_2}$ for clarity. All the combinatorial and numerical/coupling factors are in the bracket in the first line. Note that for the Maldacena vertex, we have not taken the non-local term \cite{Maldacena_2003}, as that term is 0 (or rather subleading) in the soft limit. After putting $\bm{k_1}=\bm{-k_2}$ and simplifying we get:

\begin{equation}
    \langle \gamma^{h_1}(\bm{k_1}) \gamma^{h_2} (\bm{k_2})\zeta(\bm{k_3})|_{k_3 \to 0}\rangle |_{RR}= -\frac{15c_1H^6}{16M^6_{pl}k_1^3k_3^3}
\end{equation}

\section{Soft limit of $\langle\gamma \zeta\zeta \rangle $ for
$^{(3)}\!R\delta K$}\label{apC}
The ``exchange diagram'' is the same as the right diagram in Figure 2 and we have, to the 1st order in $c_1$: 
\begin{flalign}
    O_{2\zeta}&=\int_{} 4 a\frac{M^2_{pl}}{M_5} \partial^2 \zeta \left(\epsilon \dot{\zeta}-\frac{\partial^2 \zeta}{H}a^{-2} \right)\\
    O_{\gamma \zeta \zeta}&= \text{Maldacena terms}~~ + 4\frac{M^2_{pl}}{M_5} \left( 2 \frac{\gamma_{ij}a^{-1}}{H} \partial_{i}\partial_{j} \zeta \partial^2 \zeta - a\gamma_{ij}\partial_i\partial_j \zeta \dot{\zeta}\right) &
\end{flalign}

This gives a correction to the 3-point function, in the limit $\bm{k_1} \to 0$
\begin{flalign}
   \langle\gamma^{h}(\bm{k_1}) \zeta(\bm{k_2})\zeta(\bm{k_3}) \rangle= -B_{LL}- B_{RR}+ B_{RL}+ B_{LR}+ B_{\text{contact}} 
\end{flalign}
where the contact vertex is the one from $\mathcal{O}_{\gamma \zeta \zeta}$. One finds that $B_{RR}+B_{LL}= B_{\text{contact}}$ and we can easily see that \begin{equation}
\begin{split}
    B_{LR}(k_1 \to 0)=B_{RL}(k_1 \to 0)= \frac{1}{2}\frac{B_{MT}(k_1\to 0,k_2 = k_3) \delta P_{\zeta}(k_3)}{P_{\zeta}(k_3)}= \frac{3}{4}e^{h}_{ij}(\bm{k_1})\frac{1}{k_2^2}k_{2i} k_{2j} P_{\gamma}(k_1) \delta P_{\zeta}(k_2)
\end{split}
\end{equation}

and hence the soft limit (Equation \ref{sl-0}) is satisfied.

\section{Purely Cubic Operators}\label{apD}
Here, we give explicit expressions for operators that contribute to the mixed correlators. Note that for these calculations, we have taken \ref{q-1}, \ref{q-2} as the quadratic actions, i.e. the Maldacena quadratic action. $\Lambda_i$'s are the UV cutoffs for each term while $c_i's$ are dimensionless quantities.  \\

\textbf{Purely Cubic Operators for $\langle \gamma \gamma\zeta \rangle$}
\begin{flalign}
    \mathcal{O}_1&= \int{}{} \sqrt {-g} \Lambda_1^2 \delta K_{ij}\delta K_{ij}\delta g^{00}= \int_{}{}  \Lambda_1^2 a^{3} \frac{1}{2H} \dot{\gamma_{ij}}\dot{\gamma_{ij}} \dot{\zeta}&&\\
    \mathcal{O}_{2}&=\int{}{} \sqrt {-g}  \Lambda_2 ^{(3)}\!R_{ij} \delta K_{ij}\delta g^{00}=\int_{}{}  \Lambda_2 a^{}\frac{1}{2H} \partial^2 \gamma_{ij} \dot{\gamma_{ij}} \dot{\zeta} &&\\
    \mathcal{O}_3 &= \int_{}{} \sqrt{-g}c_3 ^{(3)}\!R_{ij}^{(3)}\!R_{ij}\delta g^{00} =\int_{}{}c_3 a^{-1} \frac{1}{2H} \partial^2 \gamma_{ij}\partial^2 \gamma_{ij} \dot{\zeta} &&\\
    \mathcal{O}_4&=\int{}{} \sqrt {-g}c_4^{(3)}\!R_{ij} \delta K_{jk}\delta K _{ki}=\int_{}{} c_4 a^{} \left (\frac{1}{2} \partial^2 \gamma_{ij} \dot{\gamma_{ij}} \dot{\zeta}-\frac{1}{2H}a^{-2}\partial^2\gamma_{ij}\dot{\gamma_{jk}} \partial_i \partial_k \zeta \right.&&\\
    &~~~~~~~~~~~~~~~~~~~~~~~~~~\left.+\frac{1}{2}\epsilon\partial^2\gamma_{ij}\dot{\gamma_{jk}} \partial_i \partial_k \partial^{-2}\dot{\zeta}-\frac{1}{4}\dot{\gamma_{ij}}{\dot{\gamma_{jk}}}\partial_i \partial_k \zeta -\frac{1}{4}\dot{\gamma_{ij}}\dot{\gamma_{ij}}\partial^2 \zeta\right)&&
    \end{flalign}
    \begin{flalign}
    \mathcal{O}_5&=\int{}{} \sqrt {-g} \frac{1}{\Lambda_5} {^{(3)}\!R}_{ij}^{(3)}\!R_{jk}\delta K_{ki}=\int_{}{}\frac{1}{\Lambda_5 }\left( \frac{1}{4H}a^{-3} \partial^2 \gamma_{ij}\partial^2\gamma_{jk}\partial_k\partial_i \zeta \right.&&\\
    &\left.- \frac{1}{4}\epsilon a^{-1}\partial^2\gamma_{ij}\partial^2\gamma_{jk}\partial_k\partial_i\partial^{-2}\dot{\zeta}\right)+\frac{1}{2}a^{-1}\partial^2\gamma_{ij}\dot{\gamma_{ik}}\partial_j\partial_k \zeta +\frac{1}{2} a^{-1}\partial^2 \gamma_{ij}\dot{\gamma_{ij}}\partial^2 \zeta &&\\
    \mathcal{O}_6&= \int{}{} \sqrt {-g}\frac{1}{\Lambda_6}{^{(3)}\!R}_{ij}^{(3)}\!R_{ij}\delta K= \int_{}{} \frac{1}{\Lambda_6}\left( \frac{1}{4H}a^{-3} \partial^2 \gamma_{ij} \partial^2 \gamma_{ij} \partial^2 \zeta - \frac{1}{4} \epsilon a^{-1} \partial^2 \gamma_{ij}\partial^2 \gamma_{ij}\dot{\zeta} \right)&&\\
    \mathcal{O}_7&=\int_{}{}\sqrt{-g} \frac{1}{\Lambda_7} {^{(3)}\!R}_{ij}\delta K_{ij} {^{(3)}\!R}= \int_{}{}\frac{1}{\Lambda_7}a^{-1} \partial^2 \gamma_{ij} \dot{\gamma_{ij}} \partial^2 \zeta&& \\
    \mathcal{O}_8 &= \int_{}{}\sqrt{-g} \frac{1}{\Lambda_{8}^2} {^{(3)}\!R}_{ij}{^{(3)}\!R}_{jk}{^{(3)}\!R}_{ki}= \int_{}{} \frac{1}{\Lambda_8^2}\left (-\frac{3}{4} a^{-3}\partial^2 \gamma_{ij}\partial^2\gamma_{jk}\partial_k\partial_i \zeta -\frac{3}{4}a^{-3}\partial^2\gamma_{ij}\partial^2 \gamma_{ij}\partial^2 \zeta \right) &&\\
    \mathcal{O}_9 &= \int_{}{}\sqrt{-g} \frac{1}{\Lambda_9^2} {^{(3)}\!R}_{ij}{^{(3)}\!R}_{ij}{^{(3)}\!R}= \int_{}{} -\frac{1}{\Lambda_9^2} a^{-3}\partial^2 \gamma_{ij}\partial^2\gamma_{ij}\partial^2\zeta  &&\\
\mathcal{O}_{10}&=\int_{}{}\sqrt{-g}\Lambda_{10} \delta K_{ij}\delta K_{jk}\delta K_{ki}= \int_{}{} ~\frac{3}{4} a^3 \Lambda_{10} \dot{\gamma_{ij}}\dot{\gamma_{jk}}\left(\partial_{k}\partial_{i} \frac{\zeta}{H}a^{-2}-\epsilon \partial_k \partial_i \partial^{-2}\dot{\zeta} \right) &&\\
\mathcal{O}_{11}&=\int_{}{}\sqrt{-g}\Lambda_{11} \delta K_{ij}\delta K_{ji}\delta K= \int_{}{} ~\frac{3}{4} a^3 \Lambda_{11} \dot{\gamma_{ij}}\dot{\gamma_{ji}}\left(\partial ^2\frac{\zeta}{H}a^{-2}-\epsilon\dot{\zeta} \right) &&
\end{flalign}
\\[1.0cm]
\textbf{Purely Cubic operators for $\langle \gamma \zeta \zeta \rangle$}

\begin{flalign}
\mathcal{O}_1&=\int_{}\sqrt{-g}  \Lambda_1^2\delta K_{ij} \widetilde{N_i} \partial_j \delta g^{00}= \int_{}{} \Lambda_{1}^2 \left ( - a\frac{\dot{\gamma_{ij}}}{H^2} \partial_i \zeta \partial_j \dot{\zeta}+a^{3} \frac{\epsilon}{H} \dot{\gamma_{ij}} \partial_i \dot{\zeta}\partial_j \partial^{-2} \dot{\zeta} \right)&&\\
\mathcal{O}_2&=\int_{}\sqrt{-g}  \Lambda_2 {^{(3)}\!R}_{ij} \delta K_{ij}\delta g^{00}= \int_{}{} \Lambda_2 \left( -a^{-1}\frac{\partial^2 \gamma_{ij}}{H^2} \partial_i\partial_j \zeta \dot{\zeta}+\epsilon a\frac{\partial^2 \gamma_{ij}}{H^2} \partial_i\partial_j \dot{\zeta} \dot{\zeta} \right) &&\\
\mathcal{O}_3&= \int_{}{} \sqrt{-g}\Lambda_3 \delta K_{ij} \partial_i \delta g^{00} \partial_j \delta g^{00}= \int_{} 2\Lambda_3 a \dot{\gamma_{ij}} \partial_i \dot{\zeta} \partial_j \dot{\zeta}&&\\
\mathcal{O}_4&=\int_{}{} \sqrt{-g} \Lambda_4 {^{(3)}\!R}_{ij} \widetilde{N_i} \partial_j \delta g^{00}=\int_{}{} \Lambda_4  \left( a^{-1}\frac{\partial^2 \gamma_{ij}}{H^2} \partial_i \zeta \partial_j \dot{\zeta}-\epsilon a\frac{\partial^2\gamma_{ij}}{H} \partial_i \dot{\zeta} \partial_j \partial^{-2}\dot{\zeta}  \right)&&\\
\mathcal{O}_5&=\int_{}{} \sqrt{-g}c_5{^{(3)}\!R}_{ij}\delta K_{ij}\delta K= \int_{}{} c_5 \left (-a^{-3}\frac{\partial^2\gamma_{ij}}{2H^2} \partial_i \partial_j \zeta\partial^2 \zeta + \epsilon a^{-1}\frac{\partial^2\gamma_{ij}}{2H} \partial^2 \zeta \partial_i \partial_j \partial^{-2} \dot{\zeta} \right.&&\\
&\left.+ ~\epsilon a^{-1}\frac{\partial^2\gamma_{ij}}{2H} \dot{\zeta} \partial_i \partial_j \zeta- \epsilon^2 a\frac{\partial^2\gamma_{ij}}{2} \dot{\zeta} \partial_i \partial_j \partial^{-2} \dot{\zeta} \right)&&\\
\mathcal{O}_6&= \int_{}{}\sqrt{-g} c_6  {^{(3)}\!R}_{ij} \delta K_{jk} \delta K_{ki}= \int_{}{}~-\frac{1}{2}c_6 a \partial^2\gamma_{ij} \partial_{j}\partial_{k}\left(-\frac{\zeta}{H}a^{-2}+\epsilon \partial^{-2}\dot{\zeta}\right)  \partial_{i}\partial_{k}\left(-\frac{\zeta}{H}a^{-2}+\epsilon \partial^{-2}\dot{\zeta}\right) &&\nonumber\\
&-c_6 a \dot{\gamma_{ij}} \partial_j \partial_k \zeta \partial_k \partial_i \left(-a^{-2} \frac{\zeta}{H}+ \epsilon \partial^{-2} \dot{\zeta}\right)-c_6 a \dot{\gamma_{ij}} \partial^2\zeta \partial_j \partial_i \left(-a^{-2} \frac{\zeta}{H}+ \epsilon \partial^{-2} \dot{\zeta}\right) &&\\
\mathcal{O}_7&= \int_{}{}\sqrt{-g} c_7 {^{(3)}\!R}_{ij}\partial_i \delta g^{00}\partial_j\delta g^{00}= \int_{}{}-2c_7 a^{-1} \frac{\partial^2 \gamma_{ij}}{H^2} \partial_i \dot{\zeta}\partial_j \dot{\zeta} &&\\
\mathcal{O}_8&= \int_{}{}\sqrt{-g}c_8 {^{(3)}\!R}_{ij} {^{(3)}\!R}_{ij}\delta g^{00}=\int_{}{}~2c_8 a^{-1} \partial^2 \gamma_{ij}\partial_i\partial_j \zeta \frac{\dot{\zeta}}{H}   &&
\end{flalign}
\begin{flalign}
\mathcal{O}_9&= \int_{}{}\sqrt{-g} \frac{1}{\Lambda_9} {^{(3)}\!R}_{ij}{^{(3)}\!R}_{jk}\delta K_{ki}= \int_{}{} \frac{1}{\Lambda_9} \left ( a^{-3}\frac{\partial^2 \gamma_{ij}}{H} \partial_i \partial_k \zeta \partial_j \partial_k \zeta + a^{-3} \partial^2 \frac{\gamma_{ij}}{H} \partial_i \partial_j \partial^2 \zeta \right. &&\\
&\left.+ \frac{1}{2} a^{-1}\dot{\gamma_{ij}} \partial_i \partial_ k \zeta \partial_j \partial_k \zeta  + a^{-1}\dot{\gamma_{ij}} \partial_i \partial_j \zeta \partial^2 \zeta -\epsilon a^{-1}  \partial^2 \gamma_{ij} \partial_j \partial_k \zeta \partial_k\partial_i \partial^{-2} \dot{\zeta}   -\epsilon a^{-1} \partial^2 \gamma_{ij} \partial^2 \zeta \partial_i\partial_j \partial^{-2} \dot{\zeta} \right)&&\\
\mathcal{O}_{10}&= \int_{}{} \sqrt{-g} \frac{1}{\Lambda_{10}}{^{(3)}\!R}_{ij} {^{(3)}\!R}_{ij}\delta K = \int _{}{} \frac{1}{\Lambda_{10}}  \left (a^{-3}\frac{\partial^2 \gamma_{ij}}{H}\partial_i \partial_j \zeta \partial^2 \zeta - \epsilon a^{-1}\partial^2 \gamma_{ij}\partial_i \partial_j \zeta \dot{ \zeta} \right) &&\\
\mathcal{O}_{11}&= \int_{}{} \frac{1}{\Lambda_{11}^2} {^{(3)}\!R}_{ij}{^{(3)}\!R}_{jk}{^{(3)}\!R}_{ki}= \int_{}{}-\frac{3}{2}\frac{1} {\Lambda_{11}^2}a^{-3} \partial^2 \gamma_{ij} \partial_i \partial _k \zeta \partial_j \partial_k \zeta- 3\frac{c_{11}} {\Lambda^2}a^{-3} \partial^2 \gamma_{ij} \partial_i \partial _j \zeta \partial^2 \zeta &&\\
\mathcal{O}_{12}&=\int_{}{}\sqrt{-g}\frac{1}{\Lambda_{12}^2}{^{(3)}\!R}_{ij}{^{(3)}\!R}_{ij}{^{(3)}\!R}= \int_{}{} -4\frac{1}{\Lambda_{12}^2}a^{-3} \partial^2 \gamma_{ij} \partial_i \partial_j \zeta \partial^2 \zeta &&\\
\mathcal{O}_{13}&=\int_{}{}\sqrt{-g}\Lambda_{13} \delta K_{ij}\delta K_{jk}\delta K_{ki}= \int_{}{} ~\frac{3}{2}a^3 \Lambda_{13} \dot{\gamma_{ij}} \partial_{j}\partial_{k}\left( \frac{\zeta}{H}a^{-2}-\epsilon \partial^{-2}\dot{\zeta}\right)\partial_k\partial_i\left( \frac{\zeta}{H}a^{-2}-\epsilon \partial^{-2}\dot{\zeta}\right) &&\\
\mathcal{O}_{14}&=\int_{}{}\sqrt{-g}\Lambda_{14} \delta K_{ij}\delta K_{ji}\delta K= \int_{}{} a^3 \Lambda_{14} \dot{\gamma_{ij}} \partial_{j}\partial_{i}\left(\frac{\zeta}{H}a^{-2}-\epsilon \partial^{-2} \dot{\zeta}\right)\left(\partial^2\frac{\zeta}{H}a^{-2}-\epsilon \dot{\zeta}\right)&&
\end{flalign}

\section{Issue with Gauge}\label{apE}
The EFT of inflation is written after fixing the time diffs, so naturally, the operators in the EFT do not respect time diffs. Therefore, they are NOT gauge-invariant. Consider the following operator
\begin{align}
   \sqrt{-g}\left(\delta g^{00}(t)\right)^{2}  {^{(3)}\!R(t)}
\end{align}
Naively if we just calculate the cubic $\zeta$ interactions coming from this operator, one can easily see that we get a non-zero answer in the unitary gauge, but zero in the flat gauge. Now, one can make this operator completely gauge-invariant by introducing the Stueckelberg field, $\pi$. The gauge invariant operator reads
\begin{align}
      \sqrt{-g}\left(\frac{\partial(t+\pi)}{\partial x^{\mu}}\frac{\partial(t+\pi)}{\partial x^{\nu}}g^{\mu\nu}+1\right)^{2} {^{(3)}\!R(t+\pi)}=  \sqrt{-g}\left[(1+\dot{\pi})^{2} g^{00} + \partial_{i}\pi\partial_{j}\pi g^{ij}\right.\\ \left.+2g^{0i}\partial_{i}\pi(1+\dot{\pi})+1\right]^{2} {^{(3)}\!R(t+\pi)}
\end{align}
Since we are interested in the cubic vertex, we need the expression for $^{(3)}\!R$ only up to first order. 
\begin{align}
    ^{(3)}\!R=\partial_{k}\Gamma_{ii}^{k}-\partial_{i}\Gamma_{ik}^{k}\\
    =-a^{-2}\partial^{2}g_{ii}
\end{align}

The operator becomes
\begin{align}
    =-\left[(1+\dot{\pi})^{2}g^{00}+\partial_{i}\pi\partial_{j}\pi g^{ij}+2g^{0i}\partial_{i}\pi(1+\dot{\pi})+1\right]^{2}e^{-2\rho}\partial^{2}g_{ii}(t+\pi)
\end{align}
Let us evaluate this operator in the two gauges \cite{Maldacena_2003}\\
\textbf{Flat gauge: $\delta \phi \neq 0$}
\begin{align}
    =-6(\delta N|_{\psi=0}-\dot{\pi})^{2}H\partial^{2}\pi\\
    =-6\left(\frac{\dot{\zeta}}{H}\right)^{2}\partial^{2}\zeta
\end{align}
\\
\textbf{Co-moving gauge: $\delta \phi =0$}
\begin{align}
=-6\frac{\dot{\zeta}^{2}}{H^{2}}\partial^{2}\zeta
\end{align}
The vertex for $\zeta$ matches in the two gauges as expected. Since, this operator just contains derivatives, at leading order, this gives a vanishing contribution to the local bispectrum.

\end{appendices}

\bibliographystyle{elsarticle-num}
\bibliography{biblio}

\end{document}